\documentclass[aps,superscriptaddress,showpacs,showkeys,reprint]{revtex4-1}
\usepackage{amsmath}
\usepackage{amsfonts}
\usepackage{amsthm}
\usepackage{graphics}
\usepackage{graphicx}
\usepackage{subfigure}
\usepackage{amssymb}
\usepackage{color}
\usepackage{url}
\usepackage{hyperref}
\usepackage[abs]{overpic}
\usepackage{tikz}
\usepackage{bm}

\def\kmax{k_{\rm max}}

\begin{document}

\title{Superfluid Mutual-friction Coefficients from Vortex Dynamics in the
Two-dimensional Galerkin-truncated Gross-Pitaevskii Equation}

\author{Vishwanath Shukla}
\email{research.vishwanath@gmail.com}
\affiliation{Centre for Condensed Matter Theory, Department of Physics,
Indian Institute of Science, Bangalore 560012, India}
\author{Marc Brachet}
\email{brachet@physique.ens.fr}
\affiliation{Laboratoire de Physique Statistique de l'Ecole Normale 
Sup{\'e}rieure, \\
associ{\'e} au CNRS et aux Universit{\'e}s Paris VI et VII,
24 Rue Lhomond, 75231 Paris, France}
\author{Rahul Pandit}
\email{rahul@physics.iisc.ernet.in}
\altaffiliation[\\ also at~]{Jawaharlal Nehru Centre For Advanced
Scientific Research, Jakkur, Bangalore, India.}
\affiliation{Centre for Condensed Matter Theory, Department of Physics, 
Indian Institute of Science, Bangalore 560012, India.} 

\date{\today}
\begin{abstract}
We present algorithms for the ab-initio  determination of the temperature ($T$)
dependence of the mutual-friction coefficients $\alpha$ and $\alpha'$ and the normal-fluid
density $\rho_{\rm n}$ in the two-dimensional (2D) Galerkin-truncated Gross-Pitaevskii
system. Our algorithms enable us to determine $\alpha(T)$, even though fluctuations in 2D
are considerably larger than they are in 3D. We also examine the implications of our
measurements of $\alpha'(T)$ for the Iordanskii force, whose existence is often
questioned.
\end{abstract}
\maketitle

The elucidation of the statistical properties of superfluid turbulence
and the comparison of these with their fluid-turbulence analogs is a problem
of central importance that lies at the interface between fluid dynamics
and statistical mechanics~\cite{vinen2000classical,skrbek2012developed}. 
Theoretical treatments of superfluid
turbulence use a variety of models~\cite{berloff2014modeling,barenghiroche2014PNAS,
proukakis2008FTbec}, which are
applicable at different length scales and for different interaction
strengths. At low temperatures $T$ and for weakly interacting bosons, the
Gross-Pitaevskii (GP) equation provides a good hydrodynamical description
of a superfluid with quantum vortices. If we consider
length scales that are larger than the mean separation between quantum
vortices, and if we concentrate on low-Mach-number flows, then the
two-fluid model~\cite{donnellybook,khalatnikov} of Hall, Vinen, Bekharevich, and Khalatnikov
(HVBK) provides a good description of superfluid turbulence. In the HVBK
equations, the normal and superfluid velocities are coupled by
two mutual-friction coefficients, $\alpha$ and $\alpha'$.
The determination of $\alpha$ and $\alpha'$, along with the normal-fluid density $\rho_{\rm n}$, as
functions of $T$, from (a) experiments~\cite{donnellybook,barenghi1983mfricreview,
donnelly1998omfdata}, (b) kinetic models~\cite{griffinbook,proukakis2008FTbec,berloff2014modeling}, 
or (c) the Galerkin-truncated GP equation~\cite{girogio2011longPRE,vmrnjp13} is a
challenging problem. Such studies have been carried out only in three dimensions (3D). 
Given that (a) two-dimensional (2D) and 3D fluid turbulence are qualitatively
different~\cite{boffetta2012review,pramanareview} and (b) 2D and 3D superfluids
are also qualitatively different~\cite{Kogut1979rmp,minnhagenrmp}, 
it behooves us to carry out GP-based investigations of $\alpha$ and $\alpha'$ for a 2D superfluid.

We present the first calculation of $\alpha(T), \alpha'(T)$, and
$\rho_{\rm n}(T)$ in the 2D Galerkin-truncated GP system.
The determination of $\alpha(T), \alpha'(T)$, and $\rho_{\rm n}(T)$ turns
out to be considerably more challenging in 2D than in 3D~\cite{girogio2011longPRE} 
because of large fluctuations. We obtain the
dependence of  $\alpha(T), \alpha'(T)$, and $\rho_{\rm n}(T)$ on $T$ by
using an algorithm, which allows us to examine the evolution of vortical configurations, such
as, a pair of vortices and a quadruplet of vortices, placed initially at
the corners of a square. We find that $\alpha'(T)$ is smaller than $\alpha(T)$ 
in magnitude, but nonzero;
this suggests that the Iordanskii force~\cite{Thouless:1996p3049,Volovik:1996p3055,
Wexler:1997p3057,Hall:1998p3040,Sonin:1998p3058,Wexler:1998p3041,Fuchs:1998p3051}, 
whose existence has often been questioned, does not vanish.

The Galerkin-truncated GP equation for the complex, classical field 
$\psi(\mathbf{x},t)$ of a weakly interacting 2D Bose gas is
\begin{equation}\label{eq:tgpedimless} i\frac{\partial
\psi(\mathbf{x},t)}{\partial t} = \mathcal{P}_G\Bigl[
-\alpha_0\nabla^2\psi(\mathbf{x},t) +
g\mathcal{P}_G[|\psi|^2]\psi(\mathbf{x},t)\Bigr], 
\end{equation}
where $g$ is the effective interaction strength, the Galerkin projector
$\mathcal{P}_G[\hat{\psi}(k)]=\theta(k_{\rm max}-k)\hat{\psi} (k)$, with
$\hat{\psi}$ the Fourier transform of $\psi$ and $\theta(\cdot)$ the
Heaviside function. 
This truncated GP equation (TGPE) conserves the total energy
$H=2\alpha_0\int_{\mathcal{A}}d^2x(\alpha_0|\nabla\psi|^2 +
\frac{g}{2}[\mathcal{P}_G|\psi|^2]^2)$, the total number of particles
$N=\int_{\mathcal{A}}d^2x|\psi|^2$, and the momentum
$\mathbf{P}=\alpha_0\int_{\mathcal{A}}d^2x(\psi\nabla\psi^*-\psi^*\nabla\psi)$.
The Madelung transformation
$\psi=\sqrt{\rho(\mathbf{x},t)}\exp(i\phi)$, where $\rho$ and $\phi$ are
the density and phase field, respectively, yields the velocity 
$\mathbf{v}=2\alpha_0\nabla\phi$, with the quantum of circulation $4\pi\alpha_0$, 
the sound velocity $c=\sqrt{g\rho^*}$, the healing length
$\xi=\sqrt{2\alpha_0^2/(g\rho^*)}$,  the total density
$\rho^*=N/\mathcal{A}$, and $\mathcal{A}=L^2$ is the area of our 2D,
periodic, computational domain of side $L=2\pi$ (see the Supplemental Material~\cite{supplement}
for units). 
We use the $2/3$ dealiasing rule in our pseudospectral direct numerical
simulation (DNS) of the TGPE, with the maximum wave number
$k_{max}=2/3\times N_c/2$, where $N_c^2$ is the number of collocation
points~\cite{girogio2011longPRE}. This scheme ensures global-momentum conservation in our DNSs
and it is essential for capturing accurately the interactions of the normal
fluid with the superfluid vortices~\cite{girogio2011longPRE}. We use a fourth-order,
Runge-Kutta scheme, with time step $\Delta t$, for time marching.

Generic initial conditions evolve slowly, under the 2D TGPE dynamics, towards
equilibrium in the microcanonical ensemble~\cite{vmrnjp13}:
the system goes through initial transients, then
displays the onset of thermalization, which is followed by a regime of
partial thermalization, and then complete thermalization, with a low-$T$
Berezinskii-Kosterlitz-Thouless (BKT) phase, a high-$T$ phase with
unbound vortices, and a transition between these phases at $T_{\rm BKT}$.
To accelerate equilibration and to have direct control over (a) $T$, for
the desired equilibrium state, and (b) states with counterflows, we use
the generalized grand canonical ensemble with the equilibrium probability
distribution $\mathbb{P}[\psi]=\Xi^{-1}\exp[-\beta(H-\mu
N-\mathbf{w}\cdot\mathbf{P})]$, where $\Xi$ is the grand partition
function, $\beta=T^{-1}$ (we set the Boltzmann constant $k_B =1$), $\mu$
the chemical potential, $\mathbf{w}=\mathbf{v}_n-\mathbf{v}_s$ the
counterflow velocity, and $\mathbf{v}_n$ and $\mathbf{v}_s$ the normal and
superfluid velocities, respectively. We construct a stochastic process, which leads to
this $\mathbb{P}[\psi]$, via the 2D stochastic Ginzburg-Landau equation (SGLE)
\begin{equation} \label{eq:sglemaintext}
\frac{\partial \psi}{\partial t} = \mathcal{P}_{G}\Bigl[\alpha_0\nabla^2\psi
-g\mathcal{P}_{G}[|\psi|^2]\psi
+ \mu\psi -i\mathbf{w}\cdot\nabla\psi 
+ \zeta(\mathbf{x},t)\Bigr],
\end{equation}
where $\zeta$ is a zero-mean, Gaussian white noise with
$\langle\zeta(\mathbf{x},t)\zeta^*(\mathbf{x}',t')\rangle =
D\delta(\mathbf{x}-\mathbf{x}')\delta(t-t')$, $\delta$ the Dirac
delta function, and $D=1/(2\alpha\beta)$, in accordance with the
fluctuation-dissipation theorem.  We solve this SGLE along with
\begin{equation}
\frac{d\mu}{dt}=-\frac{\nu_N}{\mathcal{A}}(N-N_{av}),
\end{equation}
so that $N_{av}$ controls the mean value of $N$ and $\nu_N$ governs the rate at 
which the SGLE equilibrates.  The counterflow term
$i\mathbf{w}\cdot\nabla\psi$ yields states with a
non-vanishing $\mathbf{w}$.

In the HVBK model~\cite{donnellybook,barenghi1983mfricreview,2dhvbkvar}, 
a superfluid vortex does not move with the superfluid velocity
$\mathbf{v}_{\rm s}$ but with velocity 
\begin{equation}\label{eq:vortexvelmf}
\mathbf{v} = \mathbf{v}_{\rm sl} + \alpha\mathbf{s}'\times(\mathbf{v}_{\rm
n}-\mathbf{v}_{\rm sl}) - \alpha'\mathbf{s}'\times[\mathbf{s}'\times
(\mathbf{v}_{\rm n}-\mathbf{v}_{\rm sl})],
\end{equation}
where $\mathbf{v}_{\rm sl}=\mathbf{v}_{\rm s} + \mathbf{v}_{\rm si}$ is
the local superfluid velocity, with $\mathbf{v}_{\rm s}$ and
$\mathbf{v}_{\rm si}$ the imposed superfluid velocity and the self-induced
velocity because of the vortices, respectively, and $\mathbf{s}'$ the unit tangent at a
point on the vortex, with position vector $\mathbf{s}$  
\footnote{Equation~(\ref{eq:vortexvelmf}) is normally written in three dimensions (3D).
To use it in 2D, it is simplest to use a 2D projection of an infinitely long
and straight vortical filament in 3D.}.
We use the following two initial configurations in the 2D TGPE: (1) $\psi_{\tt IC1}=\psi_{\rm
pair}\psi_{\rm eq}$; and (2) $\psi_{\tt IC2}=\psi_{\rm lattice}\psi^{\rm
cf}_{\rm eq}$.  We obtain $\psi_{\tt IC1}$ by
(a) first preparing a state $\psi_{\rm pair}$, which corresponds to a
small, vortex-antivortex pair translating with a constant velocity along
the $x$ direction (Supplemental Material~\cite{supplement}) 
and (b) then combining it with an equilibrium
state $\psi_{\rm eq}$ to include finite-temperature effects
(Supplemental Material~\cite{supplement}). 
To obtain $\psi_{\tt IC2}$, we first prepare
$\psi_{\rm lattice}$, in which we place vortices of alternating signs on
the corners of a square (a vortex lattice by virtue of the
periodic boundary conditions) (Supplemental Material~\cite{supplement}); and then we include
finite-temperature and counterflow effects by multiplying $\psi_{\rm lattice}$ with the state
$\psi^{\rm cf}_{\rm eq}$ (Supplemental Material~\cite{supplement}). We obtain $\psi_{\rm eq}$ and $\psi^{\rm
cf}_{\rm eq}$ by solving the SGLE~(\ref{eq:sglemaintext}); and then we use
$\psi_{\tt IC1}$ to determine $\alpha(T)$ and $\psi_{\tt IC2}$ to calculate
both $\alpha(T)$ and $\alpha'(T)$.

\begin{figure}
\centering
\resizebox{\linewidth}{!}{
\includegraphics[height=4.0cm,unit=1mm]{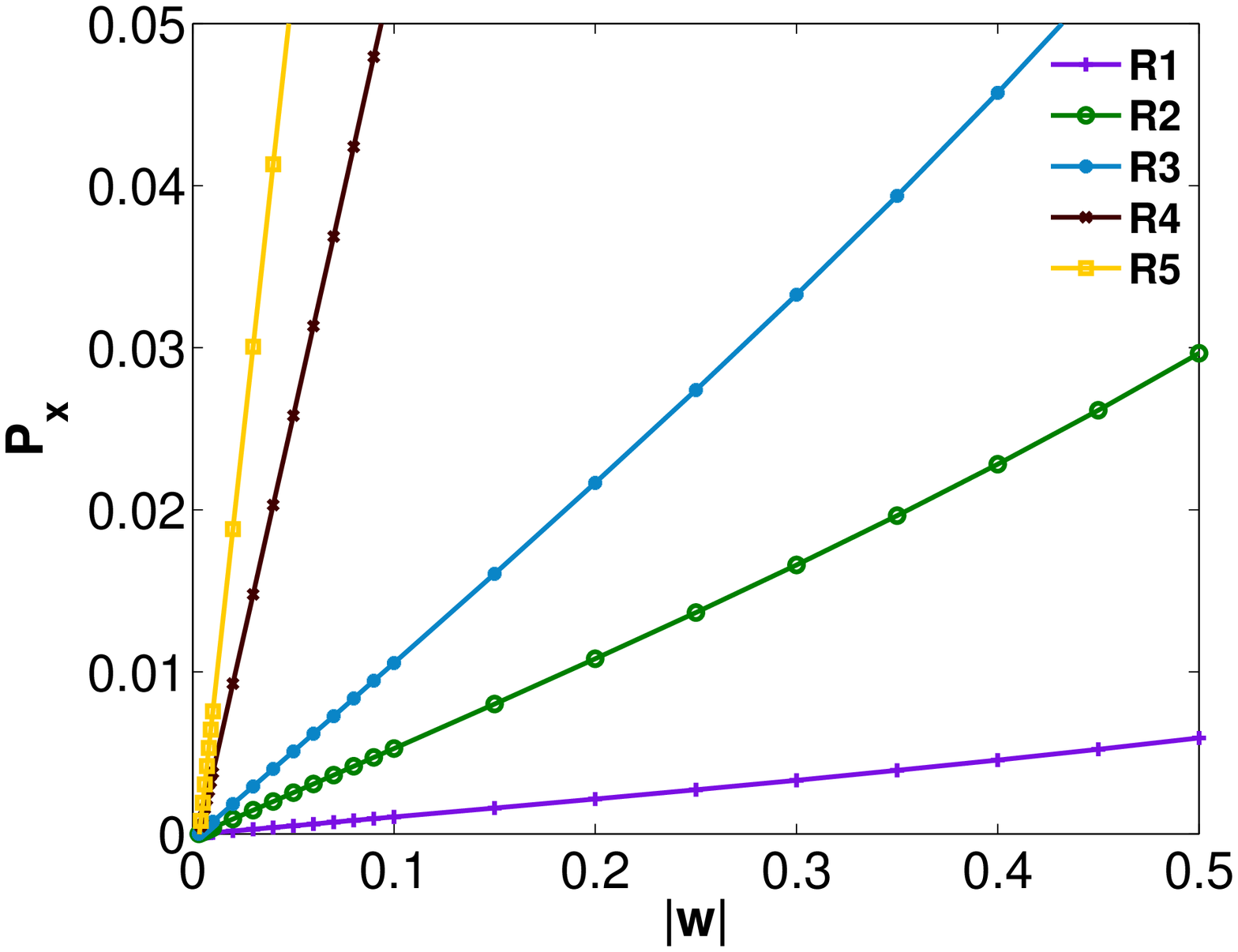}
\put(-30,30){\large{\bf (a)}}
\includegraphics[height=4.0cm,unit=1mm]{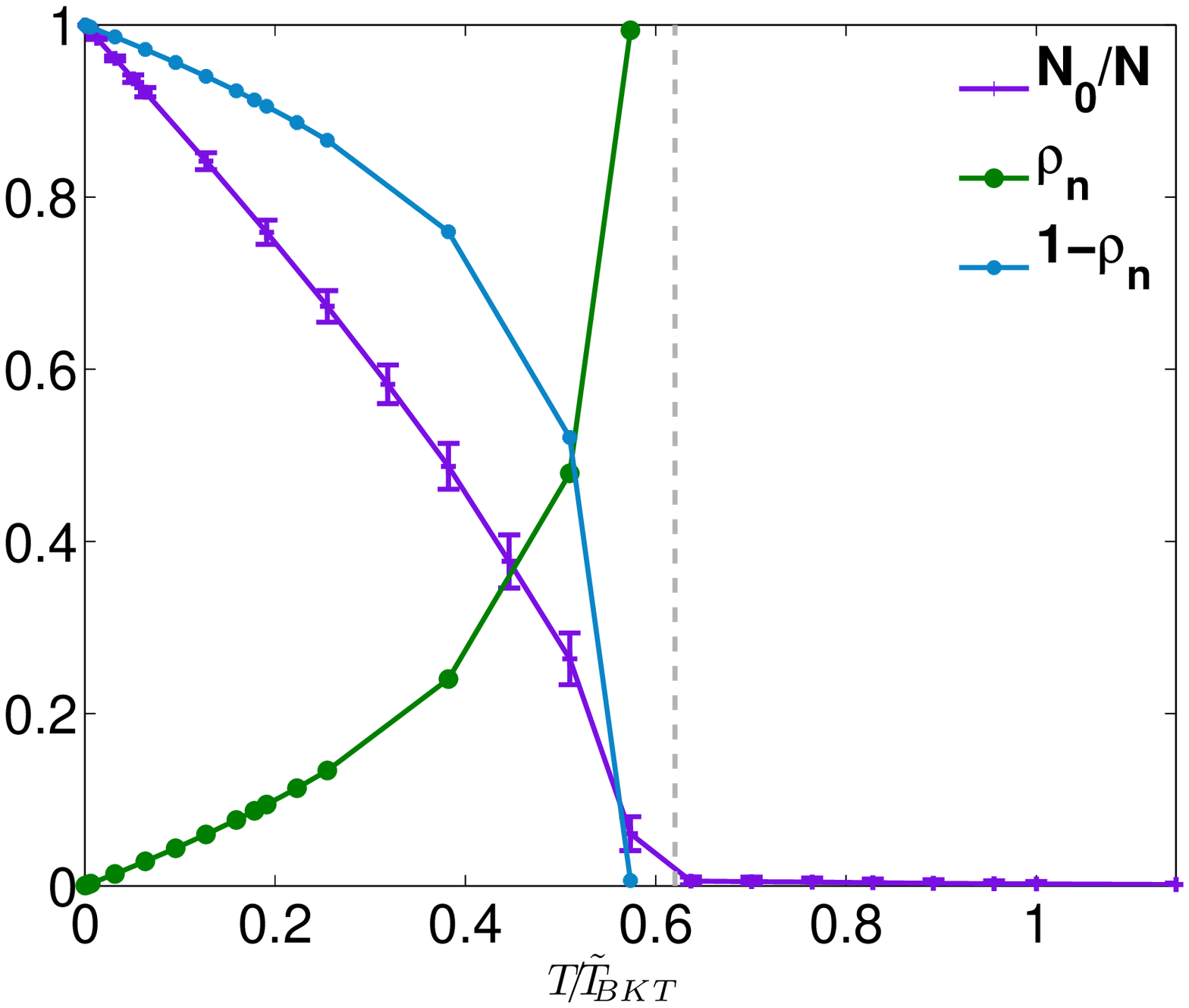}
\put(-30,30){\large{\bf (b)}}
}
\\
\resizebox{\linewidth}{!}{
\includegraphics[height=4.0cm,unit=1mm]{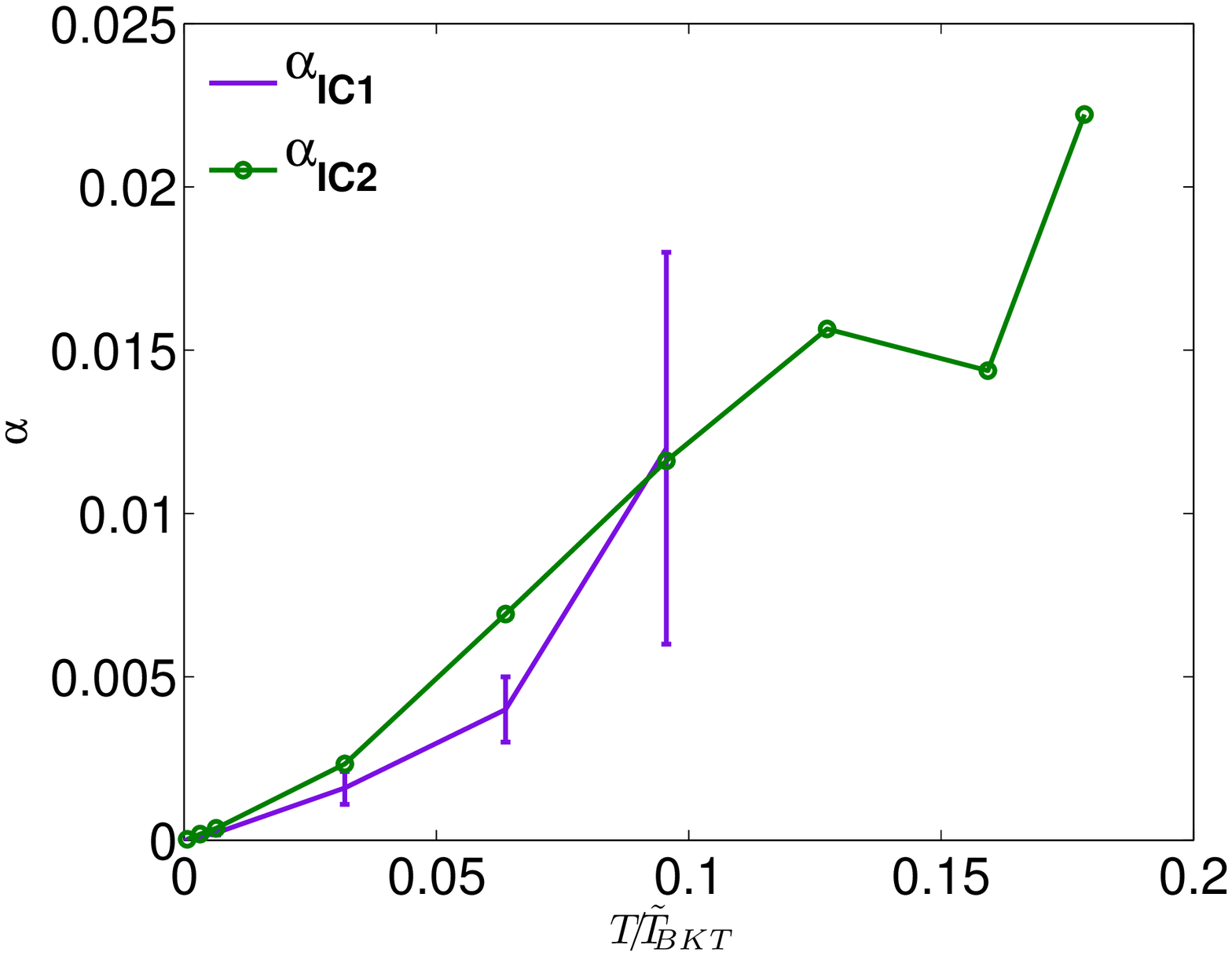}
\put(-30,30){\large{\bf (c)}}
\includegraphics[height=4.0cm,unit=1mm]{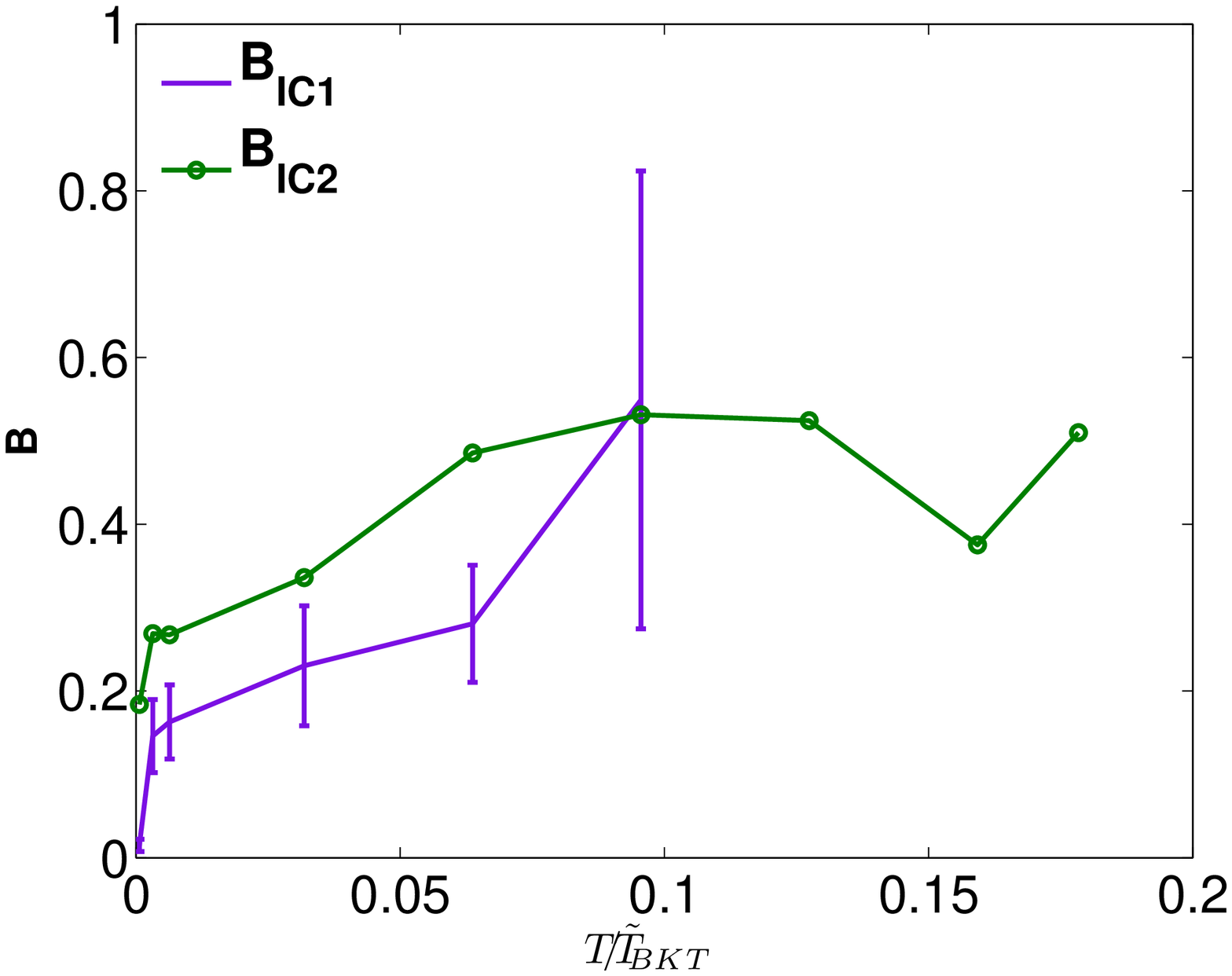}
\put(-30,30){\large{\bf (d)}}
}
\caption{(Color online) Plots of
(a) the momentum $P_{\rm x}$ versus the applied counterflow velocity $w$ for the DNS runs 
$\tt R1$-$\tt R5$;
(b) the condensate fraction $N_0/N$ (purple line), the normal fluid density $\rho_{\rm n}$
(green line), and $1-\rho_{\rm n}$ (sky-blue line) versus $T/\tilde{T}_{BKT}$;
(c) the mutual friction coefficients $\alpha_{\tt IC1}$ (purple line) and 
$\alpha_{\tt IC2}$ (green line) versus $T/\tilde{T}_{BKT}$;
(d) $B=2\alpha/\frac{\rho_n}{\rho}$ versus $T/\tilde{T}_{BKT}$.
Here the subscripts on $\alpha$ refer to the initial conditions $\tt IC1$ and $\tt IC2$.
}
\label{fig:condfrac_cf_mfcoeff}
\end{figure}

Our DNS of the TGPE Eq.~(\ref{eq:tgpedimless}) yields the spatiotemporal evolutions 
of $\psi_{\tt IC1}$ and $\psi_{\tt IC2}$. We take $\mathbf{w}=v_n\hat{x}$ for
all our SGLE DNSs with counterflows. Parameters for our DNSs are summarized in
Table~\ref{table:paramf}. We first plot the $x$ component of the momentum $P_{x}$ versus $v_n$, 
for five representative values of $T/\tilde{T}_{\rm BKT}$ 
(Fig.~\ref{fig:condfrac_cf_mfcoeff}(a)), whence we obtain 
\begin{equation}\label{eq:rhoncf}
\rho_{\rm n}(T)=\frac{1}{\mathcal{A}}
\frac{\partial P_{\rm x}}{\partial v_{\rm n}}|_{v_{\rm n}=0},
\end{equation}
whose values we list in column $3$ of Table~\ref{table:paramf}.
In Fig.~\ref{fig:condfrac_cf_mfcoeff}(b), we plot, versus the scaled
temperature $T/\tilde{T}_{\rm BKT}$, where $\tilde{T}_{\rm BKT}$ is 
a rough, energy-entropy-argument estimate of the BKT transition 
temperature~\cite{Kogut1979rmp,minnhagenrmp}, $\rho_{\rm n}$ (green curve), 
$(1-\rho_{\rm n})$ (sky-blue curve), and the 
condensate fraction $N_0/N$ (purple line), where $N_0$ is the
population of the zero-wave-number mode. 

%%%%%%%%%%%%%%%%%%%%%%%%%%%%%%%%%%%%%%%%%%%%%%%%%%%%%%%%%%%%%%%%%%%%%%%%
\begin{table}
\begin{center}
%\footnotesize 
\small
\resizebox{\linewidth}{!}{
   \begin{tabular}{@{\extracolsep{\fill}} c r r | r | c r r}
    \hline

\multicolumn{1}{c}{ $ $} & \multicolumn{1}{c}{$T/\tilde{T}_{\rm BKT}$} & 
\multicolumn{1}{c}{$\rho_{\rm n}$} & 
\multicolumn{1}{c}{$\alpha_{\tt IC1}$} 
& \multicolumn{1}{c}{$w$} & \multicolumn{1}{c}{$\alpha_{\tt IC2}$}
& \multicolumn{1}{c}{$\alpha'_{\tt IC2}$} \\
   \hline \hline

{\tt R1} & $6.37\times 10^{-4}$ & $2.7 \times10^{-4}$ &  
$(2\pm1)\times10^{-6}$ & 
$0.8$ & $2.5\times10^{-5}$ & $-2.2\times10^{-5}$\\

{\tt R2} & $3.19\times 10^{-3}$ & $1.37\times10^{-3}$ &  
$(1.0\pm.3)\times10^{-4}$ & 
$0.8$ & $1.8\times10^{-4}$ & $-1.5\times10^{-4}$\\

{\tt R3} & $6.37\times 10^{-3}$ & $2.7 \times10^{-3}$ &  
$(2.2\pm.6)\times10^{-4}$ & 
$0.6$ & $3.6\times10^{-4}$ & $-1.8\times10^{-4}$\\

{\tt R4} & $3.19\times 10^{-2}$ & $1.39\times10^{-2}$ &  
$(1.6\pm.5)\times10^{-3}$ &  
$0.4$ & $2.3\times10^{-3}$ & $-4.5\times10^{-4}$\\

{\tt R5} & $6.37\times 10^{-2}$ & $2.85\times10^{-2}$ &  
$(4\pm1)\times10^{-3}$ & 
$0.2$ & $6.9\times10^{-3}$ & $4.0\times10^{-4}$\\

{\tt R6} & $9.56\times 10^{-2}$ & $4.37\times10^{-2}$ &  
$-$ &
$0.1$ & $1.2\times10^{-2}$ & $-1.2\times10^{-3}$\\

{\tt R7} & $1.20\times 10^{-1}$ & $5.97\times10^{-2}$ &  
$(1.2\pm.6)\times10^{-2}$ & 
$0.1$ & $1.6\times10^{-2}$ & $2.9\times10^{-3}$\\

{\tt R8} & $1.59\times10^{-1}$ & $7.66\times10^{-2}$ &
$-$ &
$0.1$ & $1.4\times10^{-2}$ & $4.2\times10^{-3}$ \\

{\tt R9} & $1.78\times10^{-1}$ & $8.71\times10^{-2}$ &
$-$ &
$0.1$ & $2.2\times10^{-2}$ & $-3.5\times10^{-3}$\\

\hline
\end{tabular}
}
\end{center}
\caption{\small Mutual-friction results from our DNS runs $\tt R1$-$\tt R9$: 
$T/\tilde{T}_{\rm BKT}$ is the scaled temperature;
$\tilde{T}_{\rm BKT}=1.57\times10^{-2}$ is the energy-entropy-argument based estimate of 
the BKT transition temperature;
$\rho_{\rm n}$ is the normal-fluid density;
$\mathbf{w}=v_n\hat{x}$ is the counterflow velocity;
$\alpha$ and $\alpha'$ are the mutual friction coefficients, where the subscripts 
$\tt IC1$ and $\tt IC2$ denote the initial configurations.
In all our DNS runs, the total average density $\rho^*=N/\mathcal{A}=1$,
the total number of collocation points $N_c^2=128^2$, the
healing length $\xi=1.44\Delta x$, $\Delta x=2\pi/N_c$, the speed of sound $c=1$, and
the quantum of circulation $\alpha_0\simeq 0.05$ are kept fixed.
}
\label{table:paramf}
\end{table}
%%%%%%%%%%%%%%%%%%%%%%%%%%%%%%%%%%%%%%%%%%%%%%%%%%%%%%%%%%%%%%%%%%%%%%%%

\begin{figure}
\centering
\resizebox{\linewidth}{!}{
\begin{overpic}
[height=4.0cm,unit=1mm]{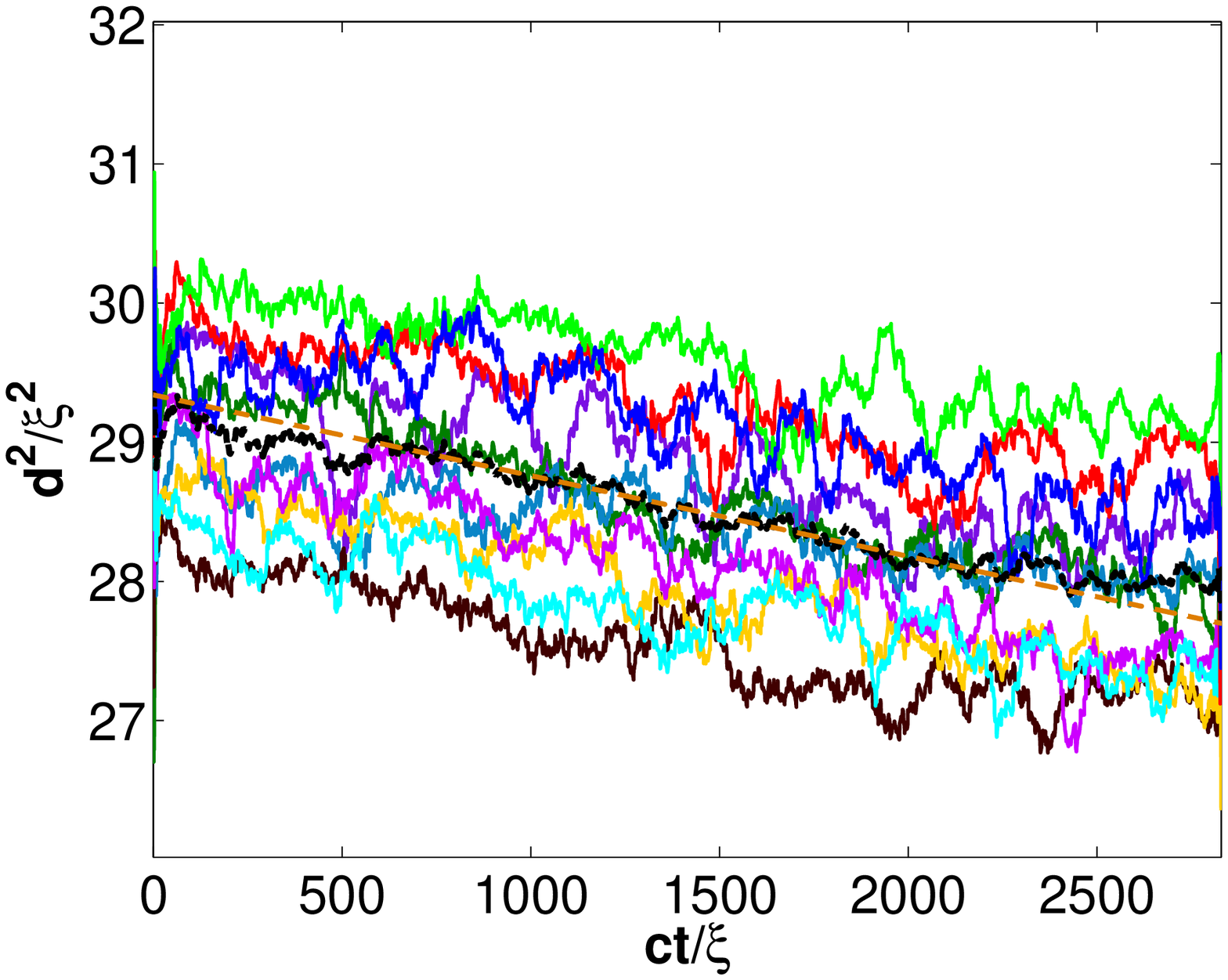}
\put(10.,10){\large{\bf (a)}}
\put(42.,35){\large{\tt R2}}
\end{overpic}
\begin{overpic}
[height=4.0cm,unit=1mm]{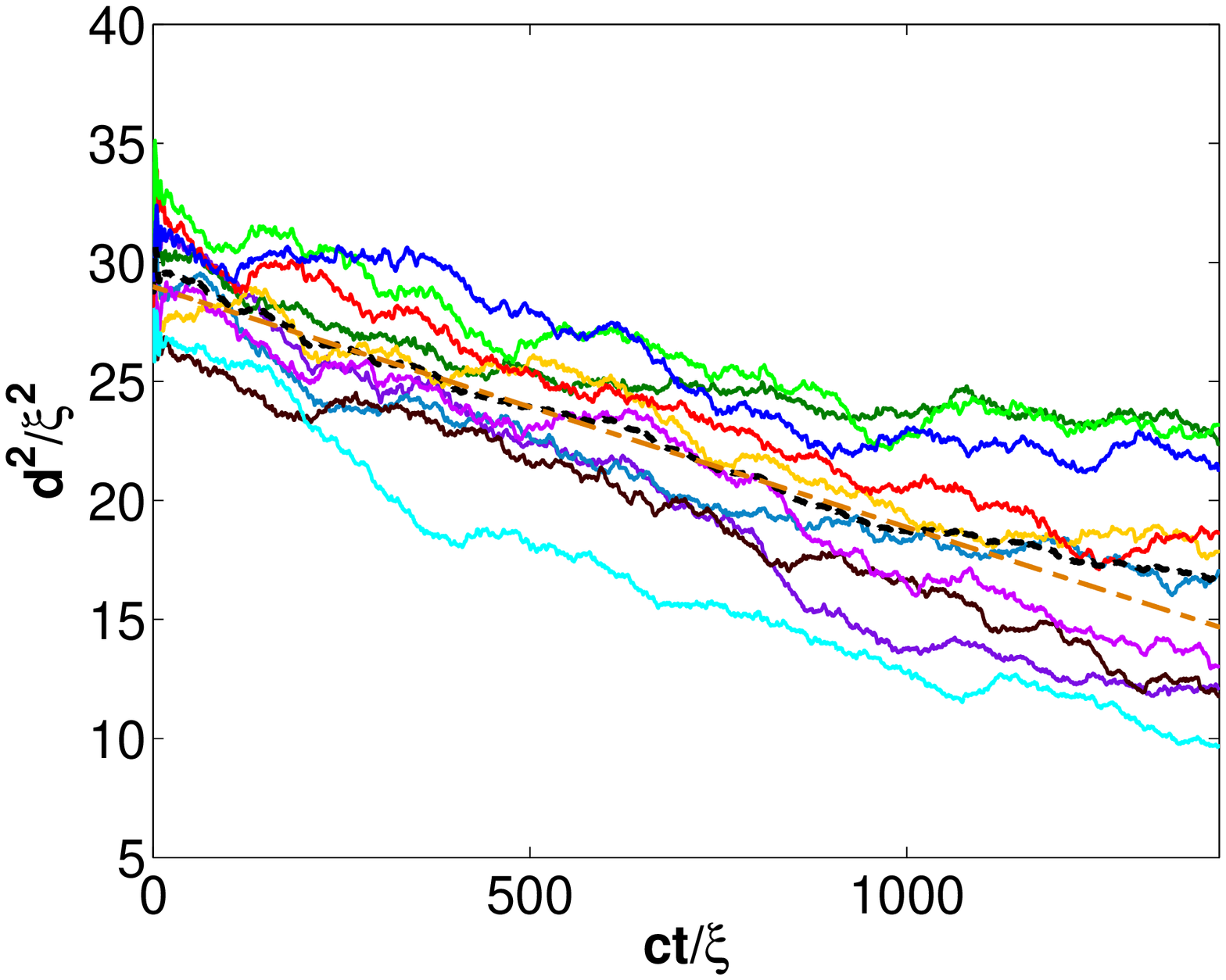}
\put(10.,10){\large{\bf (b)}}
\put(42.,35){\large{\tt R4}}
\end{overpic}
}
\caption{(Color online) Plots of the square of the vortex-pair length 
$d^2/\xi^2$ versus time $ct/\xi$ from our DNS runs:
(a) $\tt R2$ at $T/\tilde{T}_{\rm BKT}=3.19\times 10^{-3}$;
(b) $\tt R4$ at $T/\tilde{T}_{\rm BKT}=3.19\times 10^{-2}$.
For each plot, the different solid lines indicate the time evolution of $d^2/\xi^2$ for 
the different realizations of $\psi_{\rm eq}$, which we obtain from the steady state of the SGLE.
To reduce the noise in the plots of $d^2$, for these different realizations we have used a
moving-average-based smoothening procedure (the function {\textit{smooth}} in
Matlab$^\circledR$); this procedure introduces slight artifacts (high or low values of
$d^2$) near the lowest and highest values of $t$ in these plots. To obtain 
$\alpha(T)$, we use the average of all the plots of $d^2$ versus $t$ (dashed black curve),
at given value of $T$. The orange dashed line, a linear fit to this curve, is shown to
guide the eye. 
}
\label{fig:paircontraction}
\end{figure}

We begin with our results for the spatiotemporal evolution of $\psi_{\tt IC1}
=\psi_{\rm pair}\psi_{\rm eq}$. The vortex-antivortex pair in $\psi_{\rm pair}$, 
have centers that are separated, initially, by the
small distance $d(t=0)\simeq5.4\xi$ in the $y$ direction;
the pair moves at a constant velocity $v_{\rm pair}=0.2775\hat{x}$.
The state $\psi_{\rm eq}$, which is an
absolute-equilibrium state at a temperature $T<T_{\rm BKT}$, provides the
normal fluid that interacts with this vortex-antivortex pair and leads to a 
decrease in $d$ as time increases. 
In the Supplemental Material~\cite{supplement} we show that
\begin{equation}
\mathrm{d}d^2/\mathrm{d}t=-8\alpha_0(1-\alpha')\alpha;
\end{equation}
we can neglect $\alpha'$ here (as we show below, $\alpha' \ll 1$). 
Thus, we can obtain $\alpha(T)$ from the slope of a straight-line fit to a plot of 
$d^2$ versus $t$. We determine $d^2(t)$ by tracking the positions of the vortices and thus obtain 
plots such as those shown in Fig.~\ref{fig:paircontraction}
for two representative values of $T/\tilde{T}_{\rm BKT}$
(DNS runs $\tt R2$ and $\tt R4$ in Table~\ref{table:paramf}). 
The Video $\tt M1$, for the DNS run $\tt R2$, shows,
via pseudocolor plots, the spatiotemporal evolution of the field 
$|\psi(\mathbf{x},t)|^2$: the vortex-antivortex pair 
moves under the combined influence of its initial momentum and the 
finite-temperature fluctuations, the average value of $d$ decreases with time and, finally, this pair
disappears from the system (on time scales that are much longer
than those shown in this video).  From the plots in 
Fig.~\ref{fig:paircontraction} we see that (a) $d^2$ fluctuates 
significantly in time and (b) these fluctuations increase
with  $T/\tilde{T}_{\rm BKT}$ (compare Figs.~\ref{fig:paircontraction}(a) and (b)).  
Thus, the higher the temperature, the more these fluctuations limit 
our ability to determine $d^2$ reliably, with averages over a fixed
number of realizations of $\psi_{\rm eq}$, 
which we must limit, perforce, because of the computational cost of 
these calculations. We obtain $10$ values of $\alpha(T)$, 
because we use $10$ realizations of $\psi_{\rm eq}$. The mean of these values 
yield the value of $\alpha(T)$ that we have listed in column $4$ of
Table~\ref{table:paramf}; the standard deviations yield the error bars;
$\alpha$ increases with $T$ (over the range we consider). 

\begin{figure}
\centering
\resizebox{\linewidth}{!}{
\includegraphics[height=4.0cm]{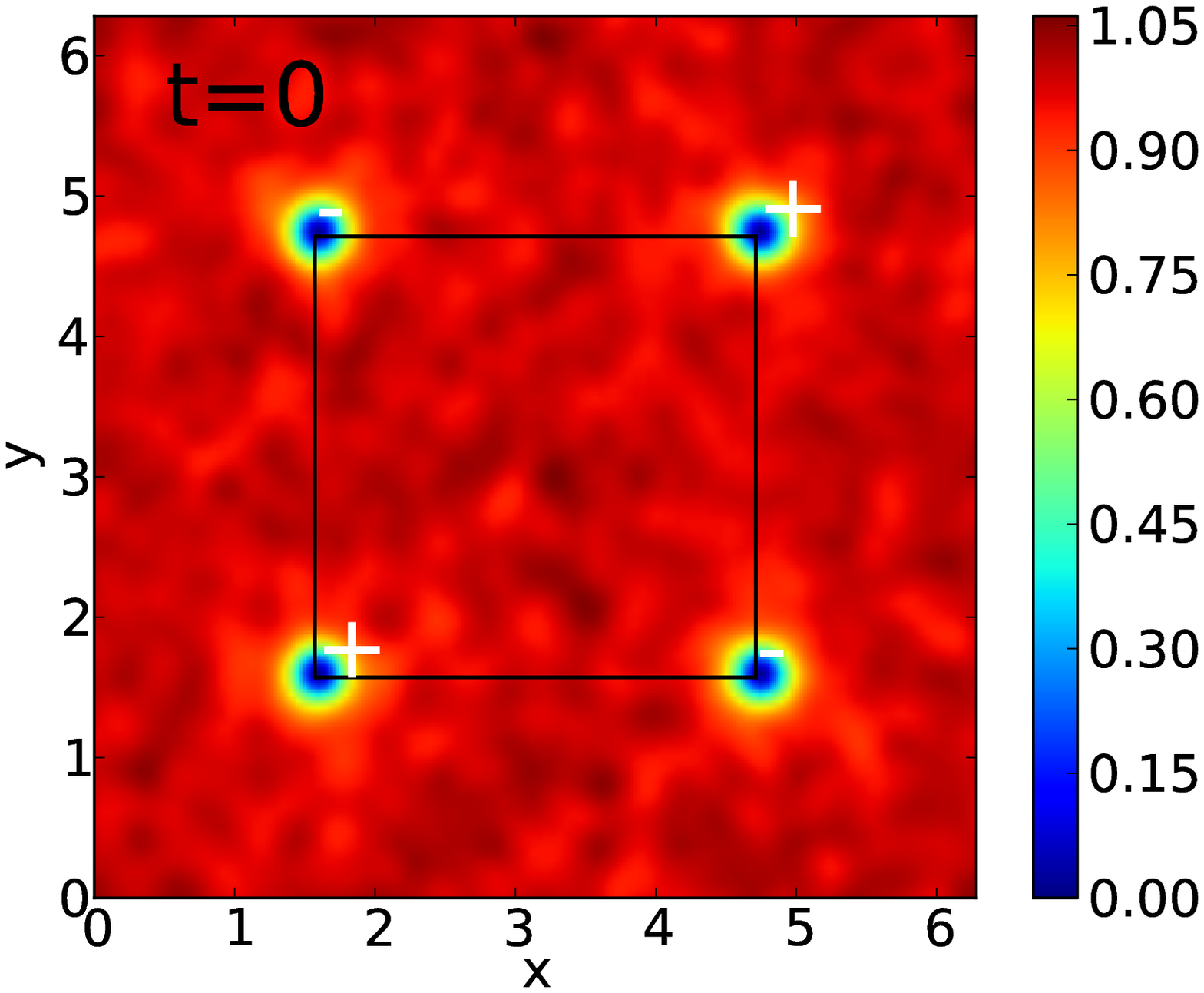}
\put(-120.,15){\large{\bf (a)}}
\includegraphics[height=4.0cm]{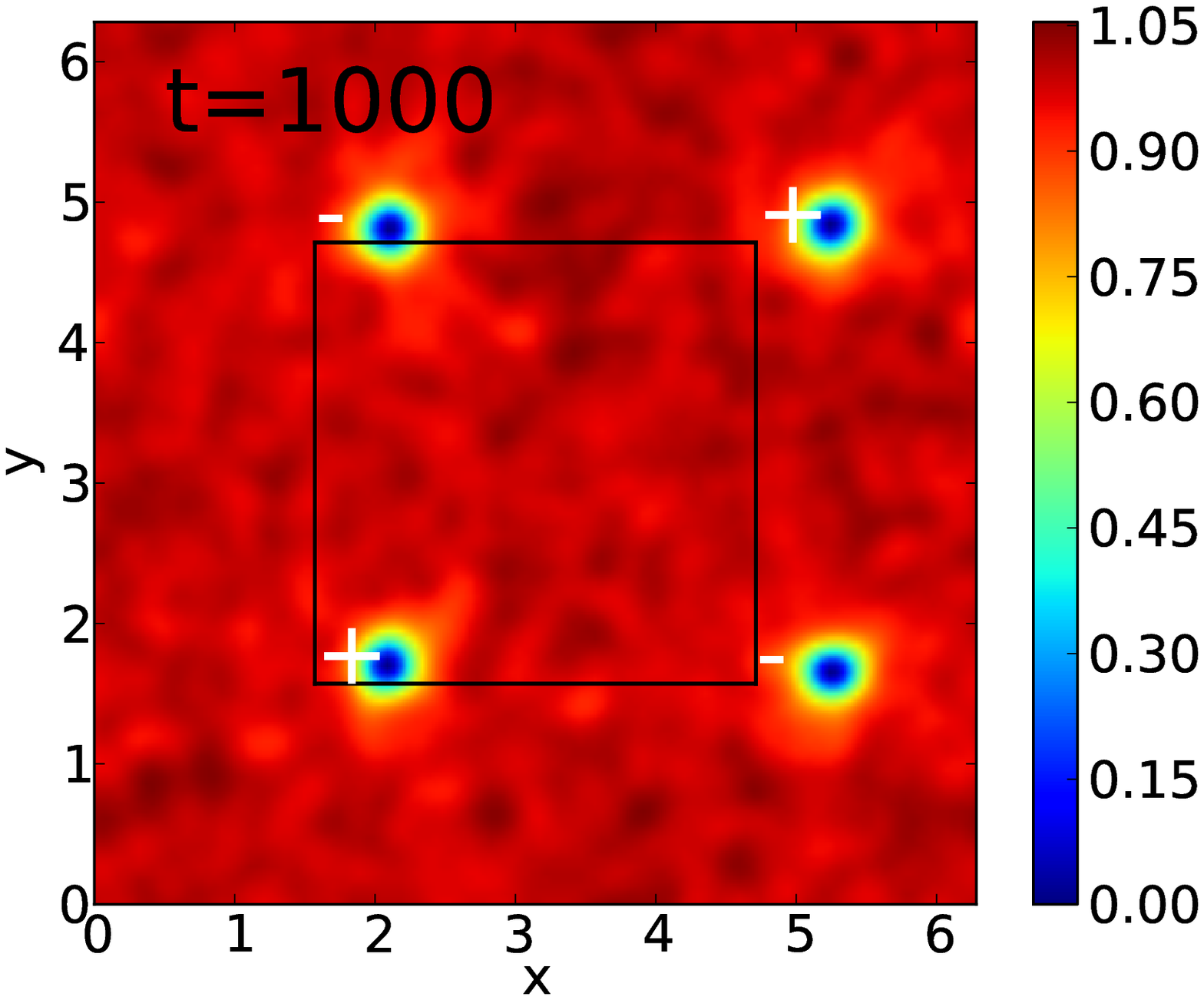}
\put(-120.,15){\large{\bf (b)}}
}
\caption{(Color online) Pseudocolor plots of the density field $|\psi(\mathbf{x},t)|^2$ from our DNS run
$\tt R2$ at two different instants of time: (a) $t=0$ and (b) $t=1000$; 
these show the drift of the vortex crystal under the imposed counterflow
$v_n=0.8 \hat{x}$ at $T/\tilde{T}_{\rm BKT}=3.19\times10^{-3}$. 
The $+$ and $-$ symbols (in white) show the signs of the vortices;
and the black frame indicates the square at whose corners we place $\pm$ vortices at
$t=0$. The Video $\tt M2$ shows, via pseudocolor plots, the spatiotemporal evolution of 
$|\psi(\mathbf{x},t)|^2$.
}
\label{fig:crystR4densitysnaps}
\end{figure}

\begin{figure}
\centering
\resizebox{\linewidth}{!}{
\includegraphics[height=4.0cm]{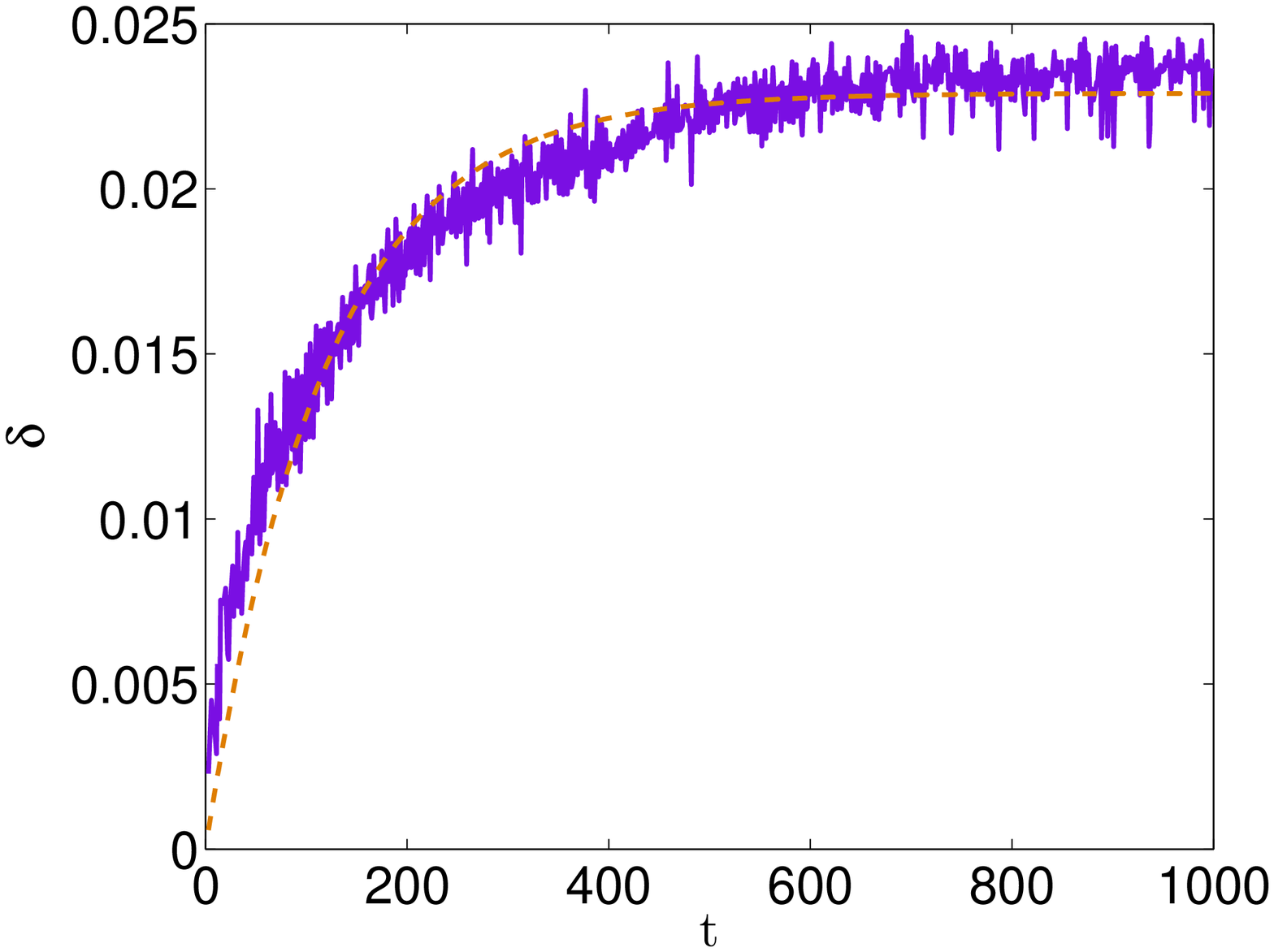}
\put(-120.,100){\large{\bf (a)}}
\put(-80,15){
\begin{tikzpicture}[scale=0.35]
\draw[thick] (0,0) rectangle (5,5);
\fill[red] (0,0.4) circle (0.3);
\node at (0,0.4) {$+$};
\node[below] at (0,-0.3) {$v_2$};
\fill[blue] (0,5-0.4) circle (0.3);
\node at (0,5-0.4) {$-$};
\node[above] at (0,5.3) {$v_3$};
\fill[blue] (5,-0.4) circle (0.3);
\node at (5,-0.4) {$-$};
\node[below] at (5-0.5,-0.3) {$v_1$};
\fill[red] (5,5.4) circle (0.3);
\node at (5,5.4) {$+$};
\node[above] at (5-0.5,5.3) {$v_4$};
%put dashed lines
\draw[thick,dashed] (0,5-.4) -- (5,5+0.4);
\draw[thick,dashed] (0,0.4) -- (5,-0.4);
\end{tikzpicture}
}
\includegraphics[height=4.0cm]{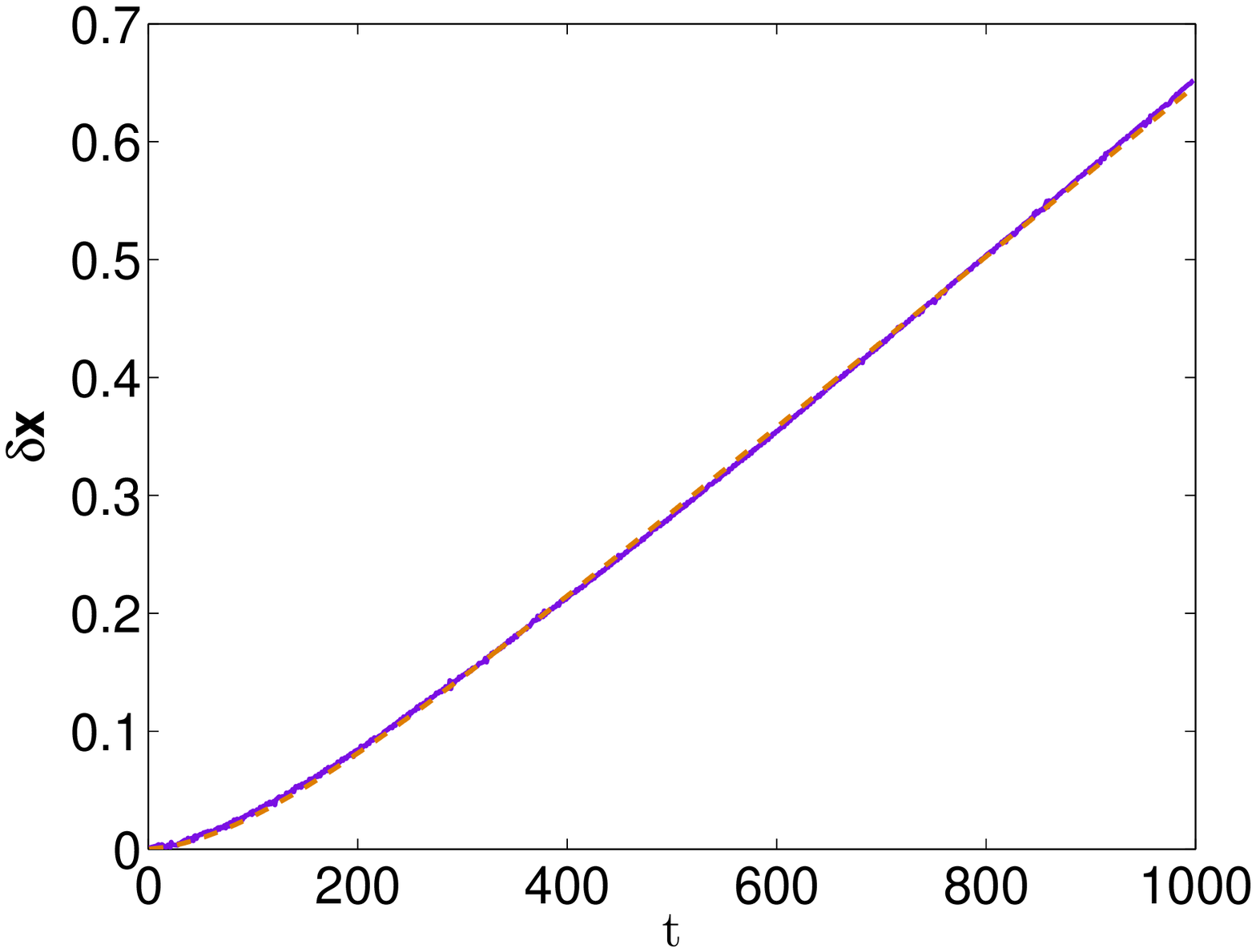}
\put(-120.,100){\large{\bf (b)}}
}
\caption{(Color online) Plots versus time $t$ of  
(a) the imperfection $\delta = \frac{1}{4}[(\delta y_2+\delta y_4)-(\delta y_1+\delta
y_3)]$ and 
(b) the drift $\delta x = \frac{1}{4}(\delta x_1 + \delta x_2 + \delta x_3 + \delta x_4)$, 
from our DNS run $\tt R2$. The orange-dashed lines indicate the fits obtained by the use
of Eqs.~(\ref{eq:ydistortmpaper}) and (\ref{eq:xdriftmpaper}) (see text).
Inset: a schematic diagram of the vortex-lattice imperfection;
the square shows the shape of the vortex lattice at $t=0$.
}
\label{fig:impdrift}
\end{figure}

The state $\psi_{\rm lattice}$ consists of a quadruplet of alternating vortices and
antivortices on the vertices of a square with sides of length $\pi$; for
this state, the self-induced velocity $\mathbf{v}_{\rm si}$, because of
these vortices and antivortices, is zero at $T=0$. In $\psi_{\tt IC2}$ we combine 
$\psi_{\rm lattice}$ with the thermalized state 
$\psi^{\rm cf}_{\rm eq}$, at different values of $T/\tilde{T}_{BKT}$ and counterflow velocity
$\mathbf{w}=v_n\hat{x}$.
Figures~\ref{fig:crystR4densitysnaps}(a) and (b) show pseudocolor plots
of the density field, for our DNS run $\tt R2$ at 
$T/\tilde{T}_{\rm BKT}=3.19\times10^{-3}$ and $v_n=0.8$, 
at two different times $t=0$ and $t=1000$. 
Figures~\ref{fig:crystR4densitysnaps}(a) and (b) and the corresponding 
Video $\tt M2$
show that the vortex lattice drifts under the influence of the imposed counterflow.
Initially the vortex-lattice has an adaptation time period, 
during which a perpendicular motion, with a negligibly small velocity, and a drift, parallel to the applied
counterflow, yield a vortex lattice imperfection, which we
quantify by $\delta = \frac{1}{4}[(\delta y_2+\delta y_4)-(\delta y_1+\delta y_3)]$, where 
$\delta y_i$ is the $y$-displacement of the vortex $i$ (see Fig.~\ref{fig:impdrift}(a)
inset); the drift parallel to the applied counterflow is given by 
$\delta x = \frac{1}{4}(\delta x_1 + \delta x_2 + \delta x_3 + \delta x_4)$, with $\delta x_i$ 
the $x$-displacement of the vortex $i$. The imperfection
$\delta(t)$ increases with $t$ and results in a self-induced 
velocity $\mathbf{v}_{\rm si}$, which leads to a decrease in the effective 
counterflow $w(t)$ because of the conservation of the total momentum. 
We develop a phenomenological model, which accounts for this effect 
(Supplemental Material~\cite{supplement}) and yields
\begin{equation}\label{eq:ydistortmpaper}
\delta(t) = \frac{\rho_nw_0}{\alpha_0(\chi_v\rho_n+\chi_p\rho)}
\biggl(1-\exp\bigl[-\frac{\alpha\alpha_0(\chi_v\rho_n+\chi_p\rho)}{\rho_n}t\bigr]\biggr)
\end{equation}
and 
\begin{equation}\label{eq:xdriftmpaper}
\begin{split}
\delta x(t) &= \biggl(\bigl[\chi_v\rho_n-\alpha'(\chi_v\rho_n+\chi_p\rho)\bigr]
\times \\
&\exp\bigl[-\frac{\alpha\alpha_0(\chi_v\rho_n+\chi_p\rho)}{\rho_n}t\bigr]
-\chi_v\rho_n \\
&+(\chi_v\rho_n+\chi_p\rho)(\alpha'+\alpha\alpha_0 t)\biggr)
\frac{\rho_nw_0}{\alpha\alpha_0(\chi_v\rho_n+\chi_p\rho)^2},
\end{split}
\end{equation}
where $w_0$ is the counterflow velocity at $t=0$; $\chi_v$ and $\chi_p$ are the
proportionality constants given by $v_{\rm si}(\delta) = \chi_v\alpha_0\delta$ and
$P_{\rm si}(\delta) = 4\pi^2\chi_P\alpha_0\rho \delta$, where $P_{\rm si}$ is the
self-induced momentum from the vortex-lattice imperfection.
We determine $\alpha(T)$ and $\alpha'(T)$ from the fits, suggested by the forms in
Eqs.~(\ref{eq:ydistortmpaper}) and (\ref{eq:xdriftmpaper}), to the plots of $\delta$ and
$\delta x$, which we obtain from our DNS runs $\tt R1$-$\tt R9$ at different temperatures
[details in the Supplemental Material~\cite{supplement}].
Figures~\ref{fig:impdrift}(a) and (b) contain plots versus $t$
of $\delta(t)$ and $\delta x(t)$, respectively; Fig.~\ref{fig:impdrift}(a) 
shows the saturation, at large $t$, of the vortex-lattice imperfection.
The values of $\alpha(T)$ and $\alpha'(T)$ that we obtain are listed in 
columns $6$ and $7$ of Table~\ref{table:paramf} for different values 
of $T/\tilde{T}_{BKT}$. It is reassuring to note that the values we obtain for $\alpha(T)$
from configurations $\tt IC1$ and $\tt IC2$ (columns $4$ and $6$) agree with each other.

The measurement of $\alpha'(T)$ in 2D is difficult because of the following 
reasons:
(1) at low temperatures its magnitude is small;
(2) at high temperatures there are large thermal fluctuations that lead to large and noisy
oscillations of the vortex lattice. Even though an accurate determination of $\alpha'(T)$ 
is difficult, we find that $\alpha'(T)$ is always nonzero and smaller in magnitude than $\alpha(T)$.
For similar studies in the 3D GPE we refer readers to 
Refs.~\cite{BerloffPRL2007dissp,girogio2011longPRE,giorgio2011PRBcf}.
In particular, Ref.~\cite{Jackson2009PRAFTvortex} has studied $\alpha(T)$ and $\alpha'(T)$
in a pancake-type condensate.

We have shown how to obtain $\alpha(T), \alpha'(T)$, and
$\rho_{\rm n}(T)$, for 2D superfluids,  by using the 2D Galerkin-truncated
GP system. Even though the determination of 
$\alpha(T), \alpha'(T)$, and $\rho_{\rm n}(T)$  is difficult,
we succeed in calculating them for
%out to be considerably more challenging in 2D than in 3D~\cite{girogio2011longPRE} 
 $T/\tilde{T}_{\rm BKT} \lesssim 10^{-1}$. At such low temperatures, the difference
between the superfluid density $\rho_s$, which should be obtained
strictly by using a helicity 
modulus~\cite{fisherbarberjasnow,Kogut1979rmp,minnhagenrmp}, and
$(1-\rho_n)$ should not be significant in typical, laboratory-scale
systems; and the HVBK model, with the values of $\alpha(T), \alpha'(T)$, 
and $\rho_{\rm n}(T)$ that we have listed in  Table~\ref{table:paramf},
should provide a good description of the dynamics of 2D superfluids
so long as we probe scales that are larger than the mean separation
between quantum vortices. 
The existence of the Iordanskii force, which is related to the third term
on the right-hand side of Eq.(\ref{eq:vortexvelmf}), has been the subject of
a debate in the latter half of the 1990s~\cite{Thouless:1996p3049,Volovik:1996p3055,
Wexler:1997p3057,Hall:1998p3040,Sonin:1998p3058,Wexler:1998p3041,Fuchs:1998p3051}.
This force is linked to the asymmetry of the scattering of quasiparticles by a
vortex~\cite{Iordanskii66a,Iordanskii66b}; and, if it is present, it implies that 
$\alpha'$ is nonzero. Thus, given that we find $\alpha' \neq 0$, our calculations imply that
there is a nonvanishing Iordanskii force. To settle conclusively the issue of the
existence of the Iordanskii force, we must obtain error bars on $\alpha'(T)$;
given the large fluctuations we have mentioned, the computational cost of obtaining such
error bars is prohibitively large.
We hope our study will lead to experimental
measurements of $\alpha(T), \alpha'(T)$, and $\rho_{\rm n}(T)$ in
2D superfluids, whose analogs for 3D superfluids~\cite{barenghi1983mfricreview}
have been known for several decades.

\vspace{0.25cm}
{\small\bf{ACKNOWLEDGEMENTS}}
\vspace{0.25cm}

We thank CSIR, UGC, DST (India) and the Indo-French Centre for Applied Mathematics (IFCAM) 
for financial support, and SERC (IISc) for computational resources. VS and RP thank ENS,
Paris for hospitality and MB thanks IISc, Bangalore for hospitality.

%%%%%%%%%%%%%%%%%%%%%%%%%%%%%%%%%%%%%%%%%%%%%%%%%%%%%%%%%
% Supplementary Material
%%%%%%%%%%%%%%%%%%%%%%%%%%%%%%%%%%%%%%%%%%%%%%%%%%%%%%%%%
\vspace{0.5cm}
{\large\bf{SUPPLEMENTAL MATERIAL}}
\vspace{0.5cm}

In this Supplemental Material we give, in Sec. I, video captions for the videos M1 and M2.
Section II is devoted to the methods we use to determine $\alpha$ and $\alpha'$ from our
DNSs. Section III contains a note on the units. Section IV describes the advective real
Ginzburg-Landau equation (ARGLE). Section V gives an overview of the
Stochastic-Ginzburg-Landau equation (SGLE) that we use. In Sections VI and VII we
generalize standard results for the low-temperature phase of the Galerkin-truncated
Gross-Pitaevskii equation by including a counterflow term. 
\section{Video Captions}

{\textcolor{blue}{\bf Video M1}}(\url{http://youtu.be/yUxRhLDeGcI}):
This video illustrates the spatiotemporal evolution of the field $|\psi(\mathbf{x},t)|^2$
for the initial configuration $\psi_{\tt IC1}=\psi_{\rm pair}\psi_{\rm eq}$
from our DNS run $\tt R2$.\newline

{\textcolor{blue}{\bf Video M2}}(\url{http://youtu.be/gMp_Rj_aMns}):
This video illustrates the spatiotemporal evolution of the field $|\psi(\mathbf{x},t)|^2$
for the initial configuration $\psi_{\tt IC2}=\psi_{\rm lattice}\psi^{\rm cf}_{\rm eq}$
from our DNS run $\tt R2$. 
%%%%%%%%%%%%%%%%%%%%%%%%%%%%%%%%%%%%%%%%%%%%%%%%%%%

\section{Mutual friction coefficients $\alpha$ and $\alpha'$}

\subsection{Determination of $\alpha$ by using the initial configuration $\tt IC1$}
\label{app:chap3app1IC1}

We can use Eq.~($4$) in the main paper %Eq.~(\ref{eq:vortexvelmf}) 
to write the distance $L_{\rm pair}(t)$ travelled in the $x$ direction 
by a vortex-antivortex pair of size $d$ (in the $y$ direction) as
\begin{equation}\label{eq:pairdisttravel}
\frac{dL_{\rm pair}}{dt} = (1-\alpha')v_{\rm si}
= (1-\alpha')\frac{\kappa}{2\pi d},
\end{equation}
where $\kappa=4\pi\alpha_0$. The time variation of $d$ is governed by
\begin{equation}\label{eq:pairsizerate}
\frac{dd}{dt} = -2\alpha\frac{dL_{\rm pair}}{dt},
\end{equation}
where $\alpha$ and $\alpha'$ are the coefficients of mutual friction. 
Equations~(\ref{eq:pairdisttravel}) and~(\ref{eq:pairsizerate}) yield 
\begin{subequations}
\begin{align}
\frac{dd}{dt} &= -4\alpha_0(1-\alpha')\alpha\frac{1}{d}; \\
\frac{dd^2}{dt} &= -8\alpha_0(1-\alpha')\alpha. \label{eq:rated2pair}
\end{align}
\end{subequations}
Therefore, 
\begin{equation}
\alpha = \frac{dd^2/dt}{8\alpha_0(1-\alpha')}
\approx \alpha = \frac{dd^2/dt}{8\alpha_0}
\end{equation}
if $\alpha'\ll1$.

%%%%%%%%%%%%%%%%%%%%%%%%%%%%%%%%%%%%%%%%%%%%%%%%%%%
\subsection{Determination of $\alpha$ and $\alpha'$ by using the initial configuration $\tt IC2$}
\label{app:chap3app1IC2}
To a first approximation, the self-induced velocity $v_{\rm si}$ and momentum $P_{\rm si}$
are linear functions of the vortex-lattice imperfection $\delta$:
\begin{equation}\label{eq:chiv}
v_{si}(\delta) = \chi_v\alpha_0\delta;
\end{equation}
and
\begin{equation}\label{eq:chip}
P_{si}(\delta) = 4\pi^2\chi_P\alpha_0\rho \delta;
\end{equation}
here $\rho$ is the total density. The coefficients $\chi_v$ and $\chi_P$ depend on the
properties of the system. We determine these by imposing a flow with velocity $v_{\rm si}$ 
on the perfect vortex lattice and then obtaining the ground state 
of this system by using the ARGLE coupled with a Newton's method 
(see Sec. IV); the vortex lattice adapts to the applied flow. 
We repeat the above procedure for different flow velocities and measure the 
imperfection $\delta$ and the momentum $P_{\rm si}$.
The coefficients $\chi_v$ and $\chi_P$ are then extracted from the slopes of the linear
fits to the plots of $v_{\rm si}$ versus $\delta$ and $P_{\rm si}$ versus $\delta$,
respectively (Eqs.(\ref{eq:chiv}) and (\ref{eq:chip})). 

From Eq.~($5$) in the main paper the counterflow momentum
\begin{equation}
P_{cf}(w) = \rho_n w\mathcal{A},
\end{equation}
where $\mathcal{A}=4\pi^2$.
Total-momentum conservation implies that an increase in the vortex-lattice
imperfection $\delta$ leads to a decrease in the effective counterflow velocity $w(\delta)$.
We have
\begin{equation}
P_0 = 4\pi^2\rho_n w_0 = P_{\rm si} + P_{cf}, 
\end{equation}
where $\rho_n$ is the normal-fluid density and $P_0$ and $w_0$ are the $t=0$ values of the
counterflow momentum and velocity, respectively. Therefore, the counterflow velocity as a function
of $\delta$ is
\begin{equation}
w(\delta) = w_0 - \frac{\chi_P\alpha_0\rho}{\rho_n}\delta.
\end{equation}
From Eq.~($4$) in the main paper the components of the velocity (for any vortex or
antivortex in our system) parallel ($\shortparallel$) and perpendicular ($\perp$) to the 
counterflow velocity are, respectively, 
\begin{equation}\label{eq:vplldelta}
\begin{split}
v^\shortparallel &= v_{\rm si}(\delta) + \alpha'[w(\delta)-v_{si}(\delta)] \\
&= \alpha_0\biggl(\chi_v - \alpha'\chi_v -\frac{\alpha'\chi_p\rho}{\rho_n}\biggr)\delta
+ \alpha'w_0
\end{split}
\end{equation}
and
\begin{equation}
\begin{split}
v^{\perp} &= \alpha[w(\delta) - v_{\rm sl}(\delta)] \\
&= \alpha w_0 - \frac{\alpha(\chi_v\rho_n+\chi_p\rho)\alpha_0}{\rho_n}\delta.
\end{split}
\end{equation}
The imperfection in the vortex lattice saturates when $v^{\perp}$ is zero,
which gives the following values for the imperfection and the drift velocity, respectively, at
saturation (subscript $\infty$):
\begin{equation}\label{eq:deltasat}
\delta_{\infty} = \frac{\rho_nw_0}{\alpha_0(\chi_v\rho_n+\chi_p\rho)};
\end{equation}
and
\begin{equation}\label{eq:vpllsat}
v^\shortparallel_{\infty}=\frac{\chi_v\rho_nw_0}{\chi_v\rho_n+\chi_p\rho}.
\end{equation}
The Eqs.~(\ref{eq:deltasat}) and (\ref{eq:vpllsat}) 
show that the large-time behavior of $\delta$ and $v^\shortparallel$ are 
independent of $\alpha'$. The equation of motion for $\delta$ is
\begin{equation}
\frac{d\delta(t)}{dt} = \alpha w_0 -
\frac{\alpha\alpha_0(\chi_v\rho_n+\chi_p\rho)}{\rho_n}\delta,
\end{equation} 
whose solution, with the initial condition $\delta(0)=0$, is 
\begin{equation}\label{eq:deltatime}
\delta(t) = \frac{\rho_nw_0}{\alpha_0(\chi_v\rho_n+\chi_p\rho)}
\biggl(1-\exp\bigl[-\frac{\alpha\alpha_0(\chi_v\rho_n+\chi_p\rho)}{\rho_n}t\bigr]\biggr).
\end{equation}
We use Eq.~(\ref{eq:deltatime}) to rewrite Eq.~(\ref{eq:vplldelta}) as
\begin{equation}
\begin{split}
v^\shortparallel(t) &= \frac{\chi_v \rho_nw_0}{\chi_v\rho_n+\chi_p\rho} + 
\biggl(\alpha'w_0 - \frac{\chi_v \rho_nw_0}{\chi_v\rho_n+\chi_p\rho}\biggr)\times \\
& \exp\bigl[-\frac{\alpha\alpha_0(\chi_v\rho_n+\chi_p\rho)}{\rho_n}t\bigr];
\end{split}
\end{equation}
the equation of motion for the drift $\delta x$ is
\begin{equation}\label{eq:eqmotiondrift}
\frac{d \delta x}{dt} = v^\shortparallel_{\rm L}(t);
\end{equation}
the solution of Eq.~(\ref{eq:eqmotiondrift}), with the initial condition $\delta
x(0)=0$, is 
\begin{equation}\label{eq:drifttime}
\begin{split}
\delta x(t) &= \biggl(\bigl[\chi_v\rho_n-\alpha'(\chi_v\rho_n+\chi_p\rho)\bigr]\times\\
&\exp\bigl[-\frac{\alpha\alpha_0(\chi_v\rho_n+\chi_p\rho)}{\rho_n}t\bigr]
-\chi_v\rho_n \\
&+(\chi_v\rho_n+\chi_p\rho)(\alpha'+\alpha\alpha_0 t)\biggr)
\frac{\rho_nw_0}{\alpha\alpha_0(\chi_v\rho_n+\chi_p\rho)^2}.
\end{split}
\end{equation}

\begin{figure*}
\centering
\resizebox{\linewidth}{!}{
\includegraphics[height=6.0cm]{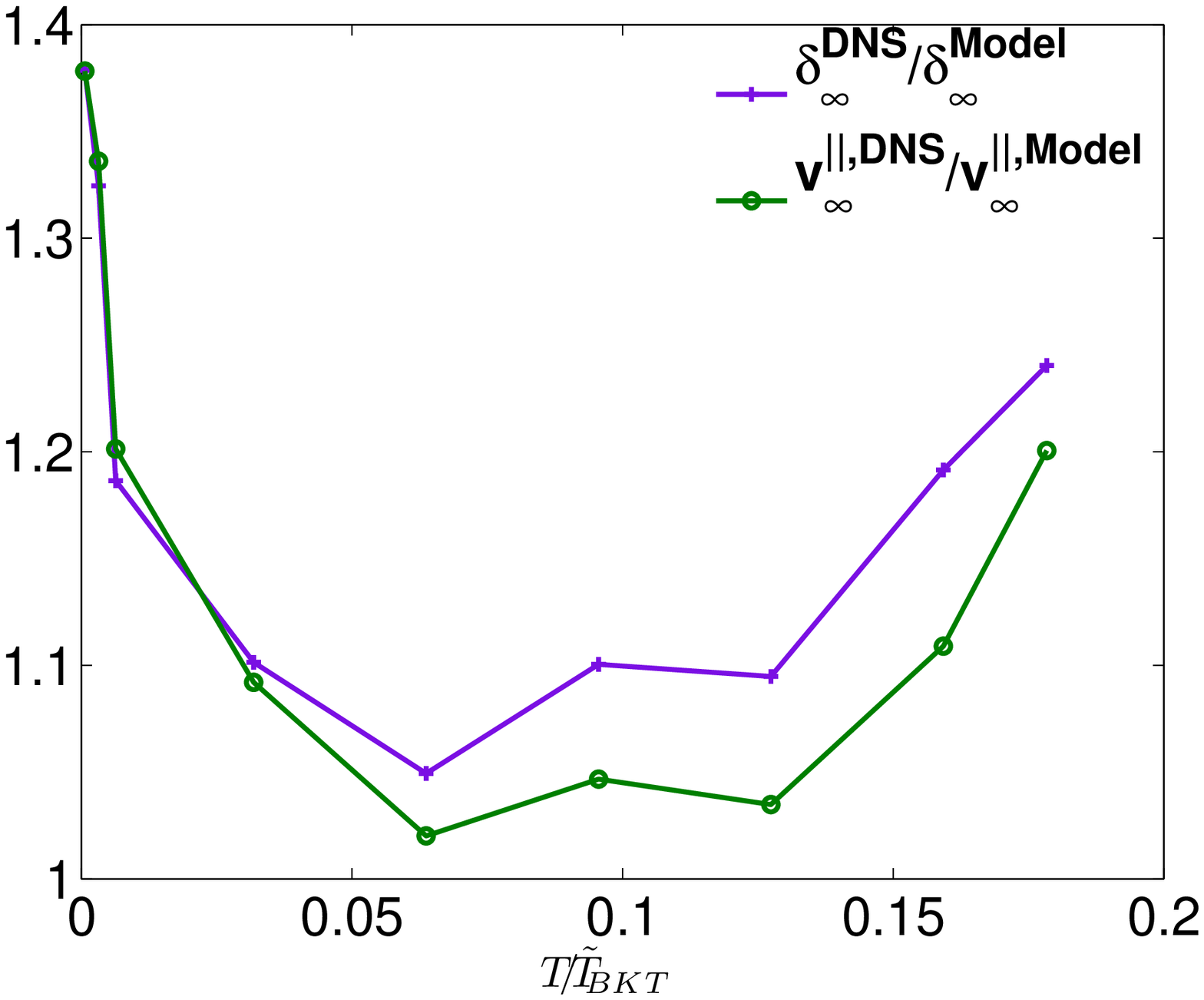}
\put(-178.,25){\large{\bf (a)}}
\includegraphics[height=6.0cm]{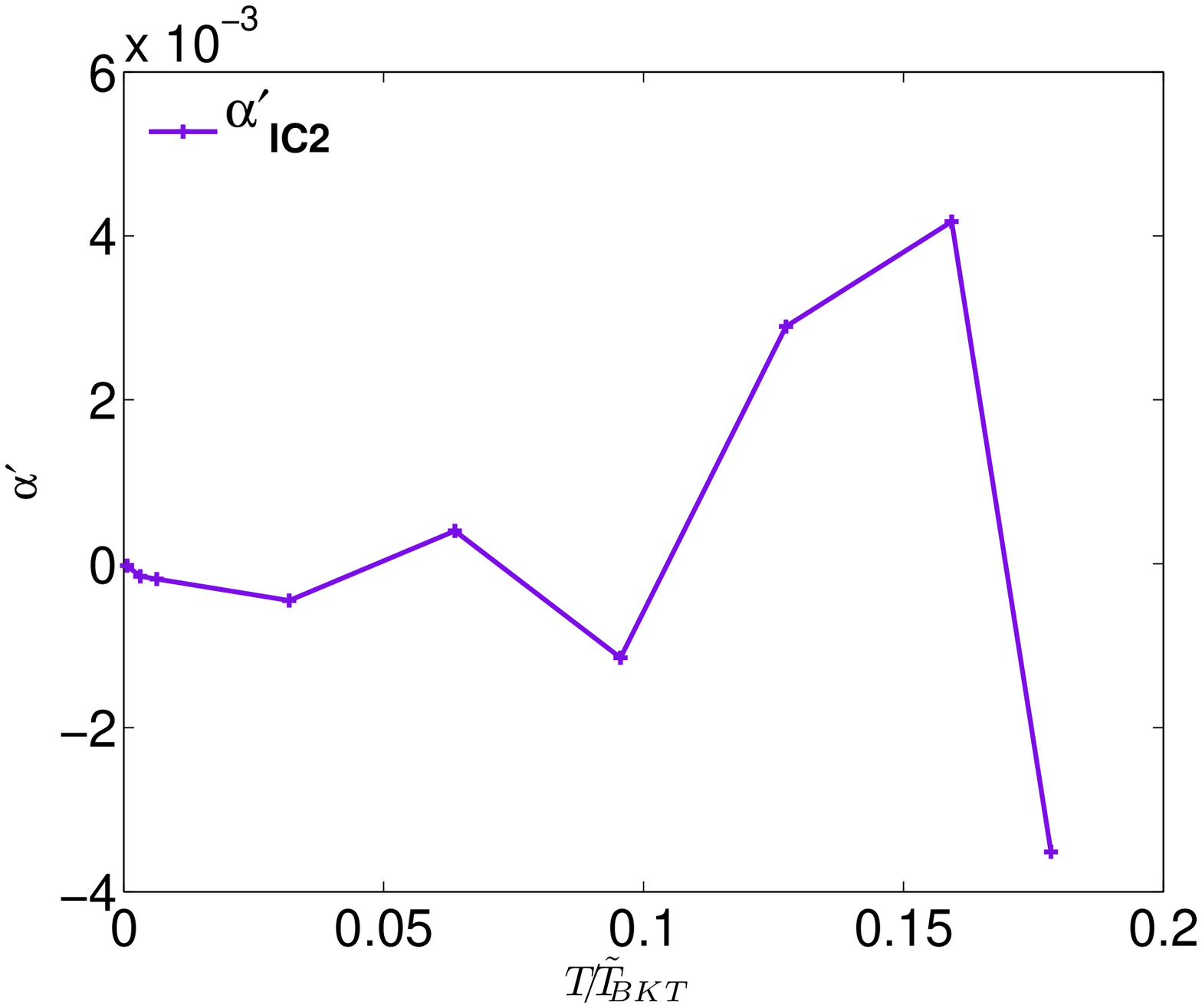}
\put(-170.,25){\large{\bf (b)}}
}
\caption{Plots of the (a) $\delta^{\rm DNS}_{\infty}/\delta^{\rm Model}_{\infty}$
and $v^{\rm\shortparallel, DNS}_{\infty}/v^{\rm\shortparallel, Model}_{\infty}$
versus $T/\tilde{T}_{BKT}$;
(b) mutual friction coefficient $\alpha'$ versus $T/\tilde{T}_{BKT}$,
obtained from the DNS runs $\tt R1$-$\tt R9$ using the initial configuration $\tt IC2$.}
\label{fig:comparednstheodeltadeltax}
\end{figure*}

To extract $\alpha$ and $\alpha'$ from our data, from the DNS runs 
$\tt R1$-$\tt R9$ with the initial configuration $\tt IC2$, we rewrite
Eqs.~(\ref{eq:deltatime}) and (\ref{eq:drifttime}), respectively, in the following 
simplified forms:
\begin{equation} \label{eq:deltatimefit}
\delta(t) = D(1-\exp(-Bt))
\end{equation}
and
\begin{equation}\label{eq:drifttimefit}
\delta x(t) = A(1-\exp(-Bt)) + Ct.
\end{equation}
The coefficients are
\begin{equation}
A = \frac{\alpha'\rho_nw_0}{\alpha\alpha_0(\chi_p\rho+\chi_v)}
-\frac{\chi_v\rho^2_nw_0}{\alpha\alpha_0(\chi_p\rho+\chi_v)^2},
\end{equation}
\begin{equation}
B = \frac{\alpha\alpha_0(\chi_p\rho+\chi_v)}{\rho_n},
\end{equation}
$C=v^\shortparallel_{\infty}$, and $D=\delta_{\infty}$. In
Fig.~\ref{fig:comparednstheodeltadeltax}(a)
we compare the values of $\delta_{\infty}$ and $v^\shortparallel_{\infty}$,
obtained from fits to our DNS data, and the predictions of our phenomenological model 
(Eqs.~(\ref{eq:deltasat}) and (\ref{eq:vpllsat})). 
Figure~\ref{fig:comparednstheodeltadeltax}(b) shows the temperature variation of 
$\alpha'$. We cannot fit this reliably to any functional form; however, we can infer that
$\alpha'$ is smaller than $\alpha$ in magnitude.

%%%%%%%%%%%%%%%%%%%%%%%%%%%%%%%%%%%%%%%%%%%%%%%%%%%

\section{Note on Units}
\label{app:chap3app2}

The GP equation, which describes the dynamical evolution of the wave function
$\psi(\mathbf{x},t)$ of a weakly interacting, 2D Bose gas at low temperatures, is
\begin{equation}\label{eq:gpech3}
i\hbar\frac{\partial \psi(\mathbf{x},t)}{\partial t}
= -\frac{\hbar^2}{2m}\nabla^2\psi(\mathbf{x},t) -\tilde{\mu}\psi(\mathbf{x},t) 
+ g_{\rm 2D}|\psi|^2\psi(\mathbf{x},t),
\end{equation}
where $g_{\rm 2D}$ is the effective interaction strength. As we have mentioned earlier,
the GP equation conserves the energy, given by the Hamiltonian
\begin{equation}
H = \int_\mathcal{A}d^2x\bigl(\frac{\hbar^2}{2m}|\nabla\psi|^2 + g_{\rm 2D}|\psi|^4\bigr),
\end{equation}
and the total number of particles $N=\int_\mathcal{A}d^2x|\psi|^2$. 
We can use the Madelung transformation to write
$\psi(\mathbf{x},t)=\sqrt{\rho(\mathbf{x},t)/m}e^{i\phi(\mathbf{x},t)}$. The total
density is $\rho^*=N/\mathcal{A}$.
To obtain Eq.($1$) in the main paper%Eq.~(\ref{eq:tgpedimless})
, we first divide Eq.~(\ref{eq:gpech3}) by $\hbar$
and define $\mu=\tilde{\mu}/\hbar$, $g=g_{\rm 2D}/\hbar$; we then set $\hbar/2m=\alpha_0$, with
$m=1$. In these units, the quantum of circulation is $h/m=4\pi\alpha_0$, the sound
velocity is $c=\sqrt{g|\psi_0|^2/m}=\sqrt{g\rho_0}$, and the healing length is
$\xi=\sqrt{\hbar^2/2m|\psi_0|^2g}=\sqrt{2\alpha_0^2/\rho_0g}$, where
$\rho_0=m|\psi_0|^2$ is the condensate density. 
%%%%%%%%%%%%%%%%%%%%%%%%%%%%%%%%%%%%%%%%%%%%%%%%%%%%%%%

\section{Advective real Ginzburg-Landau equation (ARGLE)}
\label{app:chap3appARGLE}

Compressible superfluid hydrodynamics, which is described by the GP equation, can lead, 
in the presence of vortices, to regimes dominated by acoustic emissions. 
To minimize these acoustic emissions, we prepare our initial states by using a specialized scheme, 
which we refer to as the advective-real-Ginzburg-Landau equation (ARGLE)~\cite{nore1997}. 
The desired initial states are the large-time-asymptotic solutions of the ARGLE
\begin{equation} \label{eq:argle}
\frac{\partial \psi}{\partial t} = \alpha_0\nabla^2\psi
-g|\psi|^2\psi \\ 
+ \mu\psi -i\mathbf{u}_{\rm adv}\cdot\nabla\psi 
-\frac{\mathbf{u}^2_{\rm adv}}{4\alpha_0}\psi;
\end{equation}
and these states minimize the free-energy functional
\begin{equation}
\begin{split}
\mathcal{F}_{\rm ARGLE}(\psi,\psi^*)&=\int d^3x 
\Biggl(\alpha_0\left|\nabla\psi-i\frac{\mathbf{u}_{\rm adv}}{2\alpha_0}\psi\right|^2
+\frac{1}{2}g|\psi|^4 \\
&-\mu|\psi|^2\Biggr);
\end{split}
\end{equation}
$\mathbf{u}_{\rm adv}$ is the imposed flow velocity.

\subsubsection{Numerical implementation}
We use the implicit-Euler method for time stepping in the ARGLE, i.e.,
\begin{equation}
\psi(t+\Delta t) = \frac{\psi(t) + NL(t)\Delta t}{1-L\Delta t},
\end{equation}
where we suppress the spatial argument of $\psi$,
$L=\alpha_0\nabla^2$, and $NL=(\mu-g|\psi|^2)\psi 
-i\mathbf{u}_{\rm adv}\cdot\nabla\psi -\frac{\mathbf{u}^2_{\rm adv}}{4\alpha_0}\psi$.
The field $\psi$ at the time step $(n+1)$ is given by
\begin{equation}
\hat{\psi}_{n+1} = \frac{\hat{\psi}_n+\Delta t(\mu-g\widehat{|\psi_n|^2\psi_n} 
-i\widehat{\mathbf{u}_{\rm adv}\cdot\nabla\psi_n} 
- \widehat{\frac{\mathbf{u}^2_{\rm adv}}{4\alpha_0}\psi_n})}{1-(-\alpha_0 k^2)\Delta t}.
\end{equation}
We also use Newton's method to find both the stable and the unstable fixed points  of the
above equation, which is equivalent to finding $\psi_*$, such that
\begin{equation}
F(\psi_*) \equiv \psi_*(t) - \psi_*(t+\Delta t) = 0.
\end{equation}
Every Newton step requires the solution, for $\delta \psi$, of
\begin{equation}
\frac{\delta F}{\delta \psi}\delta \psi = -F(\psi),
\end{equation}
which we obtain by an iterative bi-conjugate-gradient-stabilized method 
(BiCGSTAB)~\cite{van1992bcgstab}. This
method uses the direct application of $[\delta F/\delta\psi]$ over an arbitrary field
$\phi$, given by
\begin{equation}
\begin{split}
\frac{\delta F}{\delta \psi}\phi &= \frac{-\Delta t}{1-L\Delta t}
\Bigl[L\phi + g(2|\psi|^2\phi+\psi^2\phi^*)
-i\mathbf{u}_{\rm adv}\cdot\nabla\phi  \\
&-(\mathbf{u}^2_{\rm adv}/4\alpha_0)\phi\Bigr].
\end{split}
\end{equation}

\subsubsection{Preparation of a translating vortex-antivortex pair: $\psi_{\rm pair}$}
The steps involved in the preparation of $\psi_{\rm pair}$ are outlined below:
\begin{enumerate}
\item Initialize $\psi(x,y)=\exp(ix)$ for $l_{\rm min}<y< l_{\rm max}$ and 
$\psi(x,y)=1$ otherwise.

\item  Evolve $\psi$ by using ARGLE, with $u_{\rm adv}=0$, and allow the vortex-antivortex 
pair thus generated to contract until it reaches the desired value of the pair length $d$.

\item Evolve $\psi$, obtained in step $2$, by using ARGLE, with $\mathbf{u}_{\rm adv}=u\hat{x}$,
so that the contraction of the vortex-antivortex pair stops.

\item Use Newton's method, coupled with BiCGSTAB, to find the exact state of the
vortex-antivortex pair for $\mathbf{u}_{\rm adv}$ in step $3$ above.  
This Newton method is used to speed up the convergence to the desired solution
(because, as the pair solution is a saddle point of Eq.~(\ref{eq:argle}), the ARGLE 
procedure, if used alone, first converges, but finally ends up diverging).
\end{enumerate}

\subsubsection{Preparation of a vortex-lattice: $\psi_{\rm lattice}$}
The steps involved in the preparation of $\psi_{\rm lattice}$ are outlined below:
\begin{enumerate}

\item Initialize $\psi=\frac{(\lambda_1+\iota
\lambda_2)}{A}\tanh\bigl(\frac{A}{\sqrt{2}\xi}\bigr)$, where
$\lambda_1=\frac{1}{\sqrt{\gamma_d}}\cos x$, $\lambda_2=\frac{1}{\sqrt{\gamma_d}}\cos y$,
$\gamma_d=8/(4\pi\alpha_0)$, and $A=\sqrt{\lambda_1^2+\lambda_2^2}$.

\item Evolve $\psi$ by using ARGLE, with $u_{\rm adv,x}=\frac{1}{\gamma_d}\sin(x)\cos(y)$
and $u_{\rm adv,y}=-\frac{1}{\gamma_d}\cos(x)\sin(y)$.

\item Evolve $\psi$, obtained in step $2$, by using ARGLE, followed by Newton-BiCGSTAB, with 
$\mathbf{u}_{\rm adv}=0$  to find the exact solution.

\end{enumerate}

For more details on the preparation of an assembly of vortices, we refer
the reader to Ref.~\cite{nore1997}.

%%%%%%%%%%%%%%%%%%%%%%%%%%%%%%%%%%%%%%%%%%%%%%%%%%%
\section{Stochastic Ginzburg-Landau equation (SGLE)} \label{app:chap3appSGLE}

The stochastic Ginzburg-Landau equation (SGLE) is
\begin{equation} \label{eq:sglech3}
\frac{\partial \psi}{\partial t} = \mathcal{P}_{G}\Bigl[\alpha_0\nabla^2\psi
-g\mathcal{P}_{G}[|\psi|^2]\psi 
+ \mu\psi -i\mathbf{v}_n\cdot\nabla\psi 
+ \zeta(\mathbf{x},t)\Bigr],
\end{equation}
where $\psi$ is the wave function, $g$ the interaction strength, $\mu$ the chemical
potential, and $\mathbf{v}_n$ the counterflow velocity. $\zeta$ is a Gaussian white noise
with 
\begin{equation}
\langle\zeta(\mathbf{x},t)\rangle = 0,
\end{equation}
\begin{equation}
\langle\zeta(\mathbf{x},t)\zeta^*(\mathbf{x}',t')\rangle =
D\delta(\mathbf{x}-\mathbf{x}')\delta(t-t'),
\end{equation}
and $D=1/(2\alpha_0\beta)$, where $\beta=1/(k_{\rm B}T)$ (we set the Boltzmann constant
$k_{\rm B}=1$).

\subsubsection{Numerical implementation}

We solve the SGLE~(\ref{eq:sglech3}) along with the following, ad-hoc equation 
\begin{equation} \label{eq:musupp}
\frac{d\mu}{dt} = -\frac{\nu_N}{\mathcal{A}}(N-N_{av}),
\end{equation}
to control the number of particles $N$; the parameter $N_{av}$ controls the mean value of
$N$; and $\nu_N$ governs the rate at which the SGLE equilibrates.

The spatial Fourier-transform of Eq.~(\ref{eq:sglech3}) gives
\begin{equation} \label{eq:sgleFouriersupp}
\frac{d\hat{\psi}}{dt} = -\alpha_0 k^2\hat{\psi} -g\widehat{|\psi|^2\psi} + \mu\hat{\psi}
-i\widehat{\mathbf{v}_n\cdot\nabla\psi} + \hat{\zeta},
\end{equation}
where we have omitted the Galerkin projector $\mathcal{P}_G$ for notational simplicity.
We solve the SGLE by using a pseudospectral method with periodic boundary conditions in
space and an implicit-Euler scheme, with time step $\Delta t$, for time marching. The discrete
versions of Eqs.~(\ref{eq:musupp}) and (\ref{eq:sgleFouriersupp}) are
\begin{equation}
\mu_{n+1} = \mu_n - \Delta t\frac{\mu_N}{\mathcal{A}}(N_n-N_{av})
\end{equation} 
and
\begin{equation}
\hat{\psi}_{n+1} = \frac{\hat{\psi}_n+\Delta t(-g\widehat{|\psi_n|^2\psi_n} 
-i\widehat{\mathbf{v}_n\cdot\nabla\psi_n} )}{1+(\alpha_0 k^2-\mu_n)\Delta t} + dW_{\zeta},
\end{equation}
where $dW_{\zeta}=\sqrt{D}(d\mathcal{A})^{-1/2}\eta_{i}\sqrt{dt}$, with
$d\mathcal{A}=\Delta x\Delta y$ and $\eta_i$ are random variables that we obtain from a 
normal distribution with zero mean and unit variance.

%%%%%%%%%%%%%%%%%%%%%%%%%%%%%%%%%%%%%%%%%%%%%%%%%%%
%\chapter{Low-temperature phenomenological model}
%\label{app:chap3app3}

\section{Standard results on the BKT transition}

In this and the following Sections, we extend our discussion~\cite{vmrnjp13} of the 
low-temperature, equilibrium properties of a 2D, interacting Bose gas to situations 
in which there is a nonvanishing counterflow.

We can use the heuristic, energy-entropy 
argument to obtain a rough estimate of the BKT transition temperature 
$T_{\rm BKT}$~\cite{Kogut1979rmp,vmrnjp13}. In the $XY$ model, this transition is studied by using the 
Hamiltonian
\begin{equation} \label{eq:hamilxy:ch3}
H_{\rm{XY}}=-J \sum_{<i,j>} \cos(\theta_i-\theta_j),
\end{equation}
where $<i,j>$ denotes nearest-neighbor pairs of sites, on a 2D 
square lattice, $J$ is the nearest-neighbor exchange coupling, and
$(\theta_i - \theta_j)$ is the angle between the nearest-neighbor, 
$XY$ spins on sites $i$ and $j$. In the continuum limit, the above 
Hamiltonian becomes, to lowest order in spatial gradients, 
\begin{equation} \label{eq:conthamilxy:ch3}
H_{\rm{XY}}= \frac{J}{2} \int d^2x (\nabla \theta(x))^2.
\end{equation}
By comparing Eq.~(\ref{eq:conthamilxy:ch3}) with the kinetic-energy term 
in the energy, we find that
\begin{equation} \label{eq:coupling:ch3}
J=\frac{|\langle \psi \rangle|^2 \hbar^2}{m} =\frac{\rho \Gamma^2}{(2
\pi)^2},
\end{equation} 
where $\Gamma$ denotes the Onsager-Feynman quantum of velocity 
circulation $\Gamma=4 \pi \hbar/2 m=h/m=\kappa$. A rough estimate for the
BKT transition temperature $T_{\rm BKT}$ is given below:
\begin{equation}  \label{eq:roughTBKT:ch3}
\tilde{T}_{\rm BKT}=\frac{\pi J}{2k_B}=\frac{\pi \mid \langle \psi
\rangle \mid^2 \hbar^2}{2mk_B}
=\frac{\rho \Gamma^2}{8\pi k_B},
\end{equation}
here $\tilde{T}_{\rm BKT}$ denotes the estimate for $T_{\rm BKT}$
that follows from an energy-entropy argument~\cite{vmrnjp13}.
%%%%%%%%%

\section{Low-temperature thermodynamical computations with counterflows}

We now develop an analytical framework, which is valid at
low-temperatures $T \ll T_{\rm BKT}$, that can be used to test
some of the results of our DNS runs in the region of complete
thermalization. We calculate the equilibrium
thermodynamic functions for a weakly-interacting, 2D Bose gas,
in the grand-canonical ensemble. In the
grand-canonical ensemble the probability of a given
state is 
\begin{equation}
%\mathbb{P} = \frac{1}{\mathcal{Z}}e^{-\beta F}
\mathbb{P} = \frac{1}{\Xi}e^{-\beta (H - \mu N -\mathbf{w}\cdot \mathbf{P})},
\end{equation}
%\begin{equation}
%F = H - \mu N
%\end{equation}
where $\Xi$ is the grand partition function, $\beta$ the
inverse temperature, $\mu$ the chemical potential, $N$
the number of bosons, and $\mathbf{P}$ the momentum. The grand-canonical potential is 
\begin{equation}
\Omega = - \beta^{-1}\log(\Xi);
\end{equation}
and the mean energy $E$ and number of particles $N$ are
\begin{subequations} \label{eq:thermrel:ch3}
\begin{align}
N &= -\frac{\partial \Omega}{\partial \mu}, \\
\mathbf{P} &= -\frac{\partial \Omega}{\partial \mathbf{w}}, \\
E &= \frac{\partial \Omega}{\partial \beta} + \mu N + \mathbf{w}\cdot {\bf P}.
\end{align}
\end{subequations}
We adapt to 2D the 3D study of
Ref.~\cite{girogio2011longPRE}, expand $\psi$ in terms of 
Fourier modes $A_{\bf k}$, and obtain $\Omega$
as the sum of the saddle-point part $\Omega_{sp}$ and
$\Omega_{Q}$, the deviations from the saddle point that are
quadratic in $A_{\bf k}$. We write
$\Omega=\Omega_{sp}+\Omega_{Q}$, where
$\Omega_{sp}=-\mathcal{A}\mu^2/2g$ and  
\begin{equation} \label{eq:grandpotph:ch3}
\begin{split}
\Omega_{Q} &= - \frac{\mathcal{A}}{2\pi \beta \hbar^2} \int^{p_{\rm max}}_{0}
\Bigl(\log(\frac{2m}{\beta \sqrt{p^4+4mp^2\mu}}) \\
&+ \frac{2m^2w^2(5p^2+6m^2w^2+20m\mu)}{15(p^2+4m\mu)^2}\Bigr) p dp,
\end{split}
\end{equation}
where $\mathbf{w} = w \hat{\mathbf{x}}$.
We can also calculate the condensate depletion $\delta N$, where
the particle number $N=N_0+\delta N$, and $N_0$ is the number of particles 
in the $k=0$ mode as follows:
\begin{equation} \label{eq:deltaNintg:ch3}
\delta N = \int^{p_{\rm max}}_{0} 
\frac{mp\mathcal{A}\left(p^{-2}+\frac{1}{p^2+4m\mu}\right)}
{2\pi \beta \hbar^2} dp + \mathcal{O}(w^2).
\end{equation}
The integrals in Eqs.~(\ref{eq:grandpotph:ch3}) and (\ref{eq:deltaNintg:ch3}) 
can be performed analytically, but, in contrast to the 
3D case where the primitives are zero at $p=0$, the $2D$ primitive 
for $\Omega_{\rm ph}$ is finite at  $p=0$; and for $\delta N$ 
it is infra-red (I.R.) divergent. By subtracting the I.R. finite and 
divergent terms from $\Omega_{Q}$ and $\delta N$, respectively,
we get the following expressions, in $2D$, in the thermodynamic limit
$\mathcal{A}\to\infty$: 
\begin{equation}
\begin{split}
\Omega &= -\frac{\mu^2\mathcal{A}}{2g}
- \frac{p^2_{\rm max}\mathcal{A}}{4\pi \beta \hbar^2} 
 + \frac{m\mu \mathcal{A}\log(1+\frac{p^2_{\rm max}}{4m\mu})}
{2\pi \beta \hbar^2} \\
&-\frac{p^2_{\rm max}\mathcal{A}\log(\frac{2m}
{\beta\sqrt{p^4_{\rm max}+4m\mu p^2_{\rm max}}})}
{4\pi\beta\hbar^2}
-\frac{m^2w^2\mathcal{A}\log(1+\frac{p^2_{\rm max}}{4m\mu})}{6\pi\beta\hbar^2}\\
&+\frac{m^3w^4p^2_{\rm max}\mathcal{A}}{20\pi\mu\beta\hbar^2 p^2_{\rm max}
+80\pi m\mu^2\beta\hbar^2}
\end{split}
\end{equation}
and
\begin{equation} \label{eq:deltaN:ch3}
\delta N =m\mathcal{A} \frac{ \log(1+\frac{p^2_{\rm max}}{4m\mu}) +
\log(\frac{p^2_{\rm max}\mathcal{A}}{\hbar^2})}{4\pi \beta \hbar^2} + \mathcal{O}(w^2).
\end{equation}
By using the thermodynamic relations Eq.~(\ref{eq:thermrel:ch3}), we obtain
\begin{equation} \label{Eq:Ntot:ch3}
N=\frac{\mu \mathcal{A}}{g} - \frac{m\mathcal{A}\log(1+\frac{p^2_{\rm max}}{4m\mu})}
{2\pi \beta \hbar^2} + \mathcal{O}(w^2),
\end{equation}
\begin{equation}  \label{eq:Etot:ch3}
E=\frac{\mu^2\mathcal{A}}{2g} 
+ \frac{p^2_{\rm max}\mathcal{A}}{4\pi \beta \hbar^2} 
- \frac{m\mu \mathcal{A} \log (1 + \frac{p^2_{\rm max}}{4m\mu })}
{2\pi \beta \hbar^2} + \mathcal{O}(w^2),
\end{equation}
and
\begin{equation}  \label{eq:Px}
P_x = m^2w\mathcal{A}\frac{\log(1 + \frac{p^2_{\rm max}}{4m\mu })}{3\pi\beta\hbar^2}
+ \mathcal{O}(w^3).
\end{equation}
The expression for $P_x$ is different from the one that can be derived from the density
corresponding to the condensate depletion $mw\delta N$; this allows us to define 
$\rho_n=P_x/(w\mathcal{A})$.

\section{Low-temperature results at a given density}

We next determine the chemical potential $\mu$, which fixes the 
total density $\rho=m N/\mathcal{A}$ at a given value, by 
solving the equation  
\begin{equation} \label{eq:solvemu:ch3}
\rho - \frac{m\mu}{g} + \frac{m^2
\log (1 + \frac{p^2_{\rm max}}{4m\mu})}{2\pi \beta \hbar^2} = 0;
\end{equation}
at $\beta=\infty$, i.e., zero temperature (subscript $0$) we obtain 
\begin{equation}
\mu_0=\frac{g\,\rho}{m};
\end{equation}
to order $\beta^{-1}$ we get 
\begin{equation} \label{eq:muSol:ch3}
\mu=\mu_0+\delta \mu, 
\end{equation}
where 
\begin{equation}
\delta \mu=\frac{mg \left( 4g\rho^2 + \rho p^2_{\rm max} \right)
    \log (1 + \frac{p^2_{\rm max}}{4g \rho})}{m^2 p^2_{\rm max} + 
    2\pi \beta \hbar^2 \rho p^2_{\rm max} + 
    8\pi \beta \hbar^2 g\rho^2}.
\end{equation}
We insert $\mu$ from Eq.~(\ref{eq:muSol:ch3}) into Eq.~(\ref{eq:deltaN:ch3}), 
define the change in density $\delta \rho = m \delta N/\mathcal{A}$, 
use the energy $E$ from Eq.~(\ref{eq:Etot:ch3}), 
and then expand to order $\beta^{-1}$ to obtain
\begin{equation} \label{Eq:deltarhosol:ch3}
\delta \rho =\frac{m^2 \left( \log (1 + \frac{p^2_{\rm max}}{4g\rho}) 
+ \log (\frac{p^2_{\rm max}\mathcal{A}}{\hbar^2}) \right)}
{4\pi \beta \hbar^2},
\end{equation}
and
\begin{equation} \label{Eq:Etotsol:ch3}
E=\frac{g\rho^2\mathcal{A}}{2m^2} + \frac{p^2_{\rm max}\mathcal{A}}
   {4\pi \beta \hbar^2}. 
\end{equation}
We use Eq.~(\ref{eq:Px}) and the definition $\rho_n=P_x/(w\mathcal{A})$ to obtain
\begin{equation}
\rho_n = \frac{m^2\log(1+\frac{p^2_{\rm max}}{4g\rho})}{3\pi\beta\hbar^2}.
\end{equation}

By using Eq.~(\ref{eq:roughTBKT:ch3}) and $\rho=m \mid \langle \psi
\rangle\mid^2$, we obtain
\begin{equation} \label{Eq:betaBKTrough:ch3}
\tilde{\beta}_{\rm BKT}=\frac{1}{k_{\rm B}\tilde{T}_{\rm BKT}}
=\frac{2m^2}{\pi \rho \hbar^2}, 
\end{equation}
which we can use along with Eq.~(\ref{Eq:deltarhosol:ch3}) to relate the condensate relative 
depletion $\delta \rho /\rho$ to $\beta/\tilde{\beta}_{\rm BKT}$, 
where $\beta=1/(k_{\rm B}T)$ and $k_{\rm B}$ is the Boltzmann
constant, as given below:
\begin{equation} \label{eq:deltarhoisrhoi:ch3}
\frac{\delta \rho} {\rho} ={\frac {\tilde{\beta}_{\rm BKT}}{8\beta} } 
\log \left(\frac{p^2_{\rm max}\left(1 + \frac{p^2_{\rm max}}{4g\rho} \right) 
\mathcal{A}}{\hbar^2}\right).
\end{equation}
Similarly, the normal-fluid density fraction is
\begin{equation} \label{eq:rhonsrho}
\frac{\rho_n} {\rho} ={\frac {\tilde{\beta}_{\rm BKT}}{6\beta} } 
\log \left(1 + \frac{p^2_{\rm max}}{4g\rho} \right).
\end{equation}
We use this low-temperature result Eq.~(\ref{eq:deltarhoisrhoi:ch3})
to estimate the inverse-temperature scale $\beta_{\rm BKT}$, at which 
the depletion of the $k=0$ condensate mode becomes significant
for a finite-size system with $N_c^2$ collocation points (which fixes
the maximum momentum $p_{\rm max}$); in particular, 
we can solve Eq.~(\ref{eq:deltarhoisrhoi:ch3}), for $\delta \rho/\rho=1$, 
to obtain
\begin{equation} \label{eq:betaBKT:ch3}
{\frac {\beta_{\rm BKT}}{\tilde{\beta}_{\rm BKT}} } ={\frac {1}{8} } 
\log \left(\frac{p^2_{\rm max}\left( 1 + 
       \frac{p^2_{\rm max}}{4g\rho} \right) \mathcal{A}}{\hbar^2}\right). 
\end{equation}
By making the replacements that correspond to defining $\hbar$, 
$m$,  and $g$ in terms of $c$ and $\xi$, $p_{\rm max} \to \hbar k_{\rm max}$, 
$\hbar  \to \sqrt{2} c m \xi$, and $g \to c^2m^2/\rho$, we can
rewrite Eq.~(\ref{eq:deltarhoisrhoi:ch3}), Eq.~(\ref{eq:rhonsrho}), and
Eq.~(\ref{eq:betaBKT:ch3}) as
\begin{equation} \label{eq:Facdeltarho}
\frac{\delta \rho} {\rho}  =
{\frac {\tilde{\beta}_{\rm BKT}}{8 \, \beta} }  \log \left(\kmax^2 \mathcal{A}\,
     ( 1 + \frac{{{k_{\rm max}}}^2\,{\xi }^2}{2} ) \right),
\end{equation}
\begin{equation} \label{eq:Facrhonsurrho}
\frac{\rho_n} {\rho} ={\frac {\tilde{\beta}_{\rm BKT}}{3 \, \beta} } 
\log \left ( 1 + \frac{{{k_{\rm max}}}^2\,{\xi }^2}{2}  \right) ,
\end{equation}
and
\begin{equation} \label{eq:FacBKT:ch3}
{\frac {\beta_{\rm BKT}}{\tilde{\beta}_{\rm BKT}} } =
\frac{1}{8} \log \left(\kmax^2 \mathcal{A}\,
     ( 1 + \frac{{{k_{\rm max}}}^2\,{\xi }^2}{2} ) \right), 
\end{equation}
respectively.

\bibliographystyle{apsrev4-1}
\bibliography{references}

\end{document}